\documentclass[aps,showpacs,nofootinbib]{revtex4}

\usepackage{graphicx}  %
\usepackage{amsmath}
\usepackage{bm}  %

\newcommand{\bea}{\begin{eqnarray}}
\newcommand{\eea}{\end{eqnarray}}
\newcommand{\beq}{\begin{equation}}
\newcommand{\eeq}{\end{equation}}
\newcommand{\bqa}{\begin{eqnarray}}
\newcommand{\eqa}{\end{eqnarray}}

\newcommand{\qvec}{{\bf q}}

\def\mqo2{{\!\!\!}}

\begin{document}

\title{
Exact Relations for a Strongly-interacting Fermi Gas \\ 
near a Feshbach Resonance}

\author{Eric Braaten}\email{braaten@mps.ohio-state.edu}
\affiliation{Department of Physics,
         The Ohio State University, Columbus, OH\ 43210, USA\\}

\author{Daekyoung Kang}\email{kang@mps.ohio-state.edu}
\affiliation{Department of Physics,
         The Ohio State University, Columbus, OH\ 43210, USA\\}
\author{Lucas Platter}\email{lplatter@mps.ohio-state.edu}
\affiliation{Department of Physics,
         The Ohio State University, Columbus, OH\ 43210, USA\\}
\date{\today}
%\date{November 2007}

\begin{abstract}
A set of universal relations between various properties of any
few-body or many-body system consisting of fermions with 
two spin states and a large but finite scattering length have been 
derived by Shina Tan.
We derive generalizations of the Tan relations 
for a two-channel model for fermions near a Feshbach resonance
that includes a molecular state whose detuning energy controls
the scattering length.  We use quantum field theory methods, 
including renormalization and the operator product expansion,
to derive these relations.  They reduce to the Tan relations 
as the scattering length is made increasingly large.
\end{abstract}

\smallskip
\maketitle

%%%%%%%%%%%%%%%%%%%%%%%%%%%%%%%%%%%%%%%%%%%%%%%%%%%%%%%%%%%%%
\section{Introduction}
\label{sec:intro}
%%%%%%%%%%%%%%%%%%%%%%%%%%%%%%%%%%%%%%%%%%%%%%%%%%%%%%%%%%%%%%%%%%%%%%%%%%

Many-body systems of fermions have long been 
of great importance in astrophysics, nuclear physics, 
and solid state physics.
The development of trapping and cooling techniques for 
ultracold atoms has made them important in atomic physics as well.
In this case, the strength of the interaction is governed 
by the 2-body scattering length which can be controlled 
experimentally, adding a new dimension to the problem \cite{sps0706}.

If the scattering length $a$ is much larger than the range 
of the interactions, the system has universal properties 
that are determined only by the large scattering length.
For sufficiently low number density $n$, the universal
properties can be calculated using perturbative methods. 
If $n |a|^3$ is comparable to 1 or larger, 
the problem is nonperturbative.
In the special case of two equally-populated spin states,
systematically improvable
calculations are possible using Monte Carlo methods \cite{Lee:2008fa}.
If the populations are not equal, this approach suffers
from the fermion sign problem.
If there are three or more spin states, the problem is complicated 
by the Efimov effect \cite{Braaten:2004rn}.
The challenging nature of the general problem makes exact results 
very valuable.  One case in which exact results are known 
is the unitary limit $a = \pm \infty$, where they can be derived
by exploiting scale invariance \cite{Ho04} and conformal
invariance \cite{Son:2005rv}.

Shina Tan has derived a number of universal relations between 
various properties of an arbitrary system consisting of fermions 
in two spin states with a large scattering length
\cite{Tan0505,Tan0508,Tan0803}.
The Tan relations include 
the coefficient of the $1/k^4$ tail in the momentum distribution \cite{Tan0505}, 
a decomposition of the energy $E$ into terms that are insensitive 
to short distances \cite{Tan0505}, 
an expression for the local pair density \cite{Tan0505}, 
the rate of change of $E$ from changes in the scattering length \cite{Tan0508}, 
the relation between the pressure and the energy density 
in a homogeneous system \cite{Tan0508}, 
the virial theorem for a system in a harmonic trapping potential \cite{Tan0803},
and an expression for the inelastic 2-body loss rate \cite{Tan-private}.
These relations all involve a property of the system that Tan
called the {\it integrated contact intensity}.  For brevity, we will 
refer to it simply as the {\it contact} and denote it by $C$.
It can be expressed as the integral over space of the {\it contact density},
which we will denote by ${\cal C}$.   
The Tan relations hold for any state 
of the system: few-body or many-body, homogeneous or in a
trapping potential, superfluid or normal, zero or
nonzero temperature.
Tan derived his relations within the framework of the many-body 
Schr\"odinger equation.
He used novel methods involving generalized functions  
to implement the Bethe-Peierls boundary condition 
associated with the large 2-body scattering length.

In Ref.~\cite{Braaten:2008uh}, Braaten and Platter derived the Tan relations
using a quantum field theory formulation of the problem.
They identified the contact density ${\cal C}$ as the expectation value 
of a local composite operator constructed from the quantum fields.
To derive the Tan relations, they used standard methods of 
{\it renormalization} together with the {\it operator product expansion} (OPE).
The OPE was invented independently by Ken Wilson \cite{Wilson:1969}
and by Leo Kadanoff \cite{Kadanoff:1969} in 1969.
The OPE expresses the product of 
local operators separated by a short distance 
as an expansion in local operators with coefficients 
that are functions of the separation of the operators:
%-----------------
\begin{equation}
{\cal O}_A(\bm{R} - \mbox{$\frac12$} \bm{r}) 
{\cal O}_B(\bm{R} + \mbox{$\frac12$} \bm{r}) = 
\sum_C C_{A,B}^C(\bm{r}) {\cal O}_C (\bm{R}) .
\label{OPE-gen}
\end{equation}
%-----------------
The sum is in general an infinite sum over all possible 
local operators ${\cal O}_C (\bm{R})$.  
The functions $C_{A,B}^C(\bm{r})$ are called
{\it Wilson coefficients} or {\it short-distance coefficients}.
Some of the Wilson coefficients can be singular at $\bm{r} = 0$.
The OPE is an asymptotic expansion for small $\bm{r}$:
for any power $N$, only a finite number of terms on the right side 
of the OPE have Wilson coefficients that go to zero more slowly 
than $r^N$ as $r \to 0$ \cite{WZ:1972}.

Wilson proposed the  OPE as an alternative framework to the Lagrangian 
or Hamiltonian formulations of quantum field theory \cite{Wilson:1969}.
His motivation was the problem of the strong 
interactions associated with the nuclear force.  Some aspects 
of the strong interactions had been successfully explained 
using {\it current algebra}, which is the algebra satisfied by
the currents associated with the symmetries of a quantum field theory.
For example, if the charge operators associated with a symmetry 
satisfy the Lie algebra $[Q^a,Q^b] = i f^{abc} Q^c$, the 
associated charge density operators $\rho^a(\bm{R})$ 
satisfy the equal-time commutation relations
%-----------------
\begin{equation}
 [ \rho^a(\bm{R} - \mbox{$\frac12$} \bm{r})), 
 	\rho^b(\bm{R} + \mbox{$\frac12$} \bm{r}) ] = 
i f^{abc} \delta^3(\bm{r}) \rho^c(\bm{R}) .
\label{current}
\end{equation}
%-----------------
This is a special case of the general 
short-distance expansion in Eq.~(\ref{OPE-gen}).
Wilson observed that in a scale-invariant theory, the dependence of the
coefficients $C_{A,B}^C(\bm{r})$ on the separation $\bm{r}$
is determined up to a multiplicative constant by scale invariance
and other symmetries.  He used this observation to derive 
nontrivial results for the strong interactions that followed 
only from the assumption of scale invariance.  The quantum field theory 
that describes the strong interactions was eventually determined 
to be {\it quantum chromodynamics} (QCD).  Wilson's assumption 
of scale invariance in the strong interactions 
was justified by the discovery that the 
coupling constant of QCD is asymptotically free at short distances
\cite{Gross-Wilczek,Politzer}.  QCD and the electroweak gauge theory 
that together comprise the Standard Model of elementary particle physics 
are Lagrangian quantum field theories.  Thus the OPE has not proved to be
essential as an alternative formulation of the relativistic quantum field
theories that describe elementary particles.  However
the OPE has become a standard tool for developing systematic 
approximations to the Standard Model \cite{Collins}.  

Kadanoff used the OPE to understand 
critical phenomena in condensed matter physics \cite{Kadanoff:1969}.
He showed that the critical exponents that describe the scaling 
behavior of correlations functions at a critical point 
can be deduced from the knowledge of which Wilson coefficients 
are singular as $\bm{r} \to 0$.  
The OPE therefore provides powerful constraints on
critical behavior in condensed matter systems \cite{OPE-condmat}.
A critical point is usually 
characterized not only by scale invariance but also by 
conformal invariance.  The long-distance behavior near the 
critical point can therefore be described by a statistical 
field theory with conformal symmetry, 
i.e. a {\it conformal field theory}.  The OPE is one of the
basic tools that is used to study conformal field theories.
The quantum field theory that describes fermionic atoms 
with two spin states in the unitary limit is a 
conformal field theory.  The classification of local operators 
based on conformal symmetry has been exploited by 
Nishida and Son \cite{Nishida:2007pj}
and by Mehen \cite{Mehen:2007dn}.
However, until Ref.~\cite{Braaten:2008uh}, the OPE had not
been applied directly to the problem of
fermionic atoms with a large scattering length.

The T-matrix element for S-wave atom-atom scattering can be
written as
\beq
{\cal T}(k)=\frac{4\pi}{m}\frac{1}{k \cot\delta_0(k) -i k} ,
\label{T-k}
\eeq
where $m$ is the mass of the atoms, $k$ is the relative wavenumber, 
and $\delta_0(k)$ is the S-wave phase shift. 
If the interactions have a finite range,
the phase shift has a low energy expansion that is conveniently 
expressed in the form
\beq
k\cot \delta_0(k) = -1/a + \mbox{$\frac12$} r_s k^2 + \ldots .
\eeq
This expansion, which is called the {\it effective range expansion},
defines the scattering length $a$ and the effective range $r_s$.
The Tan relations apply to any system in which the phase shift 
at the accessible energies can be approximated by the first term 
in the effective range expansion.  We will consider the 
generalizations of these relations in two {\it scattering models}
that are specified by the S-wave phase shift.

In the {\it Zero-Range Model}, the phase shift is given by 
the leading term in the effective range expansion
up to arbitrarily high energy: 
\begin{eqnarray}
k \cot \delta_0(k) = - 1/a.
\label{delta0:ZRM}
\end{eqnarray}
This model can be formulated as a 
renormalizable local quantum field theory.  
Braaten and Platter derived the Tan relations by 
applying standard renormalization methods and the 
OPE to the quantum field theory for
the Zero-Range Model \cite{Braaten:2008uh}.  
One advantage of this derivation 
is that it can be generalized in a straightforward way to any model 
that can be formulated as a renormalizable local quantum field 
theory \cite{BKZ0709}.  One such model is the {\it Resonance Model} 
in which the S-wave phase shift is given by
\begin{eqnarray}
k \cot \delta_0(k) = - 4 \pi
\left( \lambda - \frac{g^2}{m \nu - k^2} \right)^{-1},
\label{delta0:RM}
\end{eqnarray}
where $\nu$, $g$, and $\lambda$ are real parameters.
This model was first formulated as a quantum field 
theory by Kaplan \cite{Kaplan:1996nv} 
and by Kokkelmans et al.~\cite{KMCWH02}.
It is of practical importance in cold atom physics 
because it provides a natural description of fermions 
near a Feshbach resonance.
An important special case of the Resonance Model is the 
{\it Effective Range Model}, which can be obtained by setting $\lambda=0$.
The equation for the S-wave phase shift in Eq.~(\ref{delta0:RM})
reduces in this limit to the form
\begin{equation}
k \cot \delta_0(k) = 
- 1/a + \mbox{$\frac12$} r_s k^2 ,
\label{delta0:ars}
\end{equation}
where $a$ is the scattering length and $r_s < 0$ is the effective range.
The effective range must be negative for this phase shift to arise from a
local quantum field theory that is renormalizable \cite{Phillips:1997xu}.
The Effective Range Model has been proposed by Petrov 
as a model for a narrow Feshbach resonance \cite{Petrov:2004}.

In this paper, we derive the generalizations of the Tan relations 
for the Resonance Model and the Effective Range Model.
We begin in Section~\ref{sec:Tan} by enumerating the Tan relations.
In Section~\ref{sec:ZRM}, we summarize the derivation of these 
relations in Ref.~\cite{Braaten:2008uh} using the quantum field 
theory for the Zero-Range Model.
In Section~\ref{sec:RM}, we describe the formulation of the 
Resonance Model as a renormalizable local quantum field theory.
The generalizations of the Tan relations in the Resonance Model
are derived in Sections~\ref{sec:tail}--\ref{sec:lossrate}.
We use the OPE to identify the contact 
density operator.  The generalizations of the Tan relations
involve the expectation values of two other local operators 
besides the contact density operator.  
In limits in which the scattering length becomes 
arbitrarily large, the generalized Tan relations of the 
Resonance Model must reduce to the original Tan relations.
The resulting constraints on matrix elements of local operators 
are derived in Section~\ref{sec:constraints}.
In Section~\ref{sec:ERM}, we derive the generalizations of the 
Tan relations for the Effective Range Model by exploiting 
the fact that it is a limiting case of the Resonance Model.
Our results are summarized in Section~\ref{sec:summary}.
The diagrammatic methods that we use to calculate the Wilson 
coefficients in the OPE are described in 
Appendix~\ref{app:diagrams}.

%%%%%%%%%%%%%%%%%%%%%%%%%%%%%%%%%%%%%%%%%%%%%%%%%%%%%%%%%%%%%
\section{The Tan Relations}
\label{sec:Tan}
%%%%%%%%%%%%%%%%%%%%%%%%%%%%%%%%%%%%%%%%%%%%%%%%%%%%%%%%%%%%%%%%%%%%%%%%%%

In a series of papers \cite{Tan0505,Tan0508,Tan0803},
Shina Tan derived some universal relations for states consisting of
fermions with two spin states that interact
through a large scattering length $a$ and may be trapped in
an external potential $V(\bm{R})$.  
We will refer to the fermions as atoms, 
although this problem also has applications in 
condensed matter physics and in nuclear physics. 
We label the two spin states by $\sigma = 1,2$.
The Hamiltonian can be 
expressed as the sum of a kinetic term $T$, an interaction term $I$, 
and an external potential term $V$: $H = T + I + V$.
The Tan relations all involve a quantity $C$ that we call 
the {\it contact}.
There are seven distinct Tan relations:
\begin{enumerate}

\item
{\bf Tail of the momentum distribution} \cite{Tan0505}.
The momentum distribution $\rho_{\sigma}(\bm{k})$ for atoms 
with spin $\sigma$ has a power-law tail that decreases 
at large wavenumber $\bm{k}$ like $1/k^4$:
%-----------------
\begin{equation}
\rho_{\sigma} (\bm {k}) \longrightarrow C/k^4 .
\label{tail-Tan}
\end{equation}
%-----------------
The coefficient $C$ is the same for both spin states.

\item
{\bf Energy relation} \cite{Tan0505}.
The energy $E = \langle H \rangle 
	= \langle T \rangle + \langle I \rangle + \langle V \rangle$
can be decomposed into four terms
that are each insensitive to distances much smaller than $|a|$:
%-----------------
\begin{equation}
E = 
\sum_\sigma \int \!\! \frac{d^3k}{(2 \pi)^3} \frac{k^2}{2m} 
        \bigg( \rho_{\sigma}(\bm{k}) - \frac{C}{k^4} \bigg)
+ \frac{C}{4 \pi m a} 
+ \langle V \rangle .
\label{energy-Tan}
\end{equation}
%-----------------
The interaction energy $\langle I \rangle$ has been separated into
two terms proportional to $C$.  One of them has been combined with the 
kinetic term to make the integral over $\bm{k}$ convergent.
 
\item
{\bf Local pair density} \cite{Tan0505}.
If $N_{\rm pair}(\bm{R}, s)$
is the number of pairs of atoms with spins 1 and 2
in a small ball of volume $\frac43 \pi s^3$ centered at the point
$\bm{R}$, the contact density ${\cal C}$ at that point 
can be expressed as 
%%-----------------
\begin{equation}
{\cal C}(\bm{R}) = \lim_{s\to 0}
(4/s^4) N_{\rm pair}(\bm{R}, s) .
\label{pairdensity-Tan}
\end{equation}
%-----------------
Note that $N_{\rm pair}(\bm{R}, s)$ scales as $s^4$ as $s \to 0$ 
instead of $s^6$, so the number of pairs with small separation $s$ 
is much larger than one might naively expect.

\item
{\bf Adiabatic relation} \cite{Tan0508}.
The rate of change of the energy $E =\langle H \rangle$ 
due to a change in the scattering length is
%-----------------
\begin{equation}
\frac {d \ }{da} E = \frac{\hbar^2}{4\pi m a^2} ~ C .
\label{adiabatic-Tan}
\end{equation}
%-----------------

\item
{\bf Pressure relation} \cite{Tan0508,Tan0803}.
For the homogeneous system with constant external potential,
the pressure ${\cal P}$ and the energy density ${\cal E}$ are related by
%-----------------
\begin{equation}
{\cal P} = \frac{2}{3}{\cal E} + \frac{\hbar^2}{12\pi m a} ~ {\cal C} ,
\label{pressure-Tan}
\end{equation}
%-----------------
where ${\cal C}$ is the contact density.

\item
{\bf Virial theorem} \cite{Tan0803}.
The virial theorem for atoms 
in a harmonic trapping potential
in the unitary limit 
$a = \pm \infty$ was derived in Ref.~\cite{TKT0504}.
Tan derived the virial theorem 
for the case of finite $a$ \cite{Tan0803}:
%-----------------
\begin{equation}
E = 2 \langle V \rangle - \frac{\hbar^2}{8 \pi m a} ~ C .
\label{virial-Tan}
\end{equation}
%-----------------

\item
{\bf Inelastic 2-body loss rate} \cite{Tan-private}.
If a pair of atoms has inelastic scattering channels, 
the scattering length $a$ has a negative imaginary part.  
To leading order in ${\rm Im}\, a$,
the rate at which the atoms are depleted by 
inelastic collisions is
%-----------------
\begin{equation}
\Gamma \approx \frac{\hbar^2 (-{\rm Im}\,a)}{2 \pi m |a|^2} C  .
\label{Gam-Tan}
\end{equation}
%-----------------

\end{enumerate}

The adiabatic relation in Eq.~(\ref{adiabatic-Tan})
can be used as an operational definition of the contact.
A simple example in the case $a>0$ is the shallow diatomic molecule, 
or {\it dimer}, formed by atoms with spins 1 and 2.
The energy of a dimer at rest is $E =- \hbar^2/(m a^2)$. From 
Eq.~(\ref{adiabatic-Tan}), the contact for the dimer is 
%-----------------
\begin{equation}
C = 8 \pi/a .
\label{C-dimer}
\end{equation}
%-----------------
In Ref.~\cite{Tan0508}, Tan used the adiabatic relation in 
Eq.~(\ref{adiabatic-Tan}) to determine the contact density
for the homogeneous gas consisting of equal
populations of atoms with spins 1 and 2 in three limits:
the {\it BCS limit} ($a \to 0^-$), the {\it unitary limit}
($a \to \pm \infty$), and the {\it BEC limit} ($a \to 0^+$).
If the total number density of atoms is $n$, 
the contact densities in these three limits are
%-----------------
\begin{subequations}
\begin{eqnarray}
{\cal C} &\longrightarrow& 4 \pi n/a 
\hspace{2cm} {\rm as\ } a \to 0^-,
\\
&\longrightarrow& 
\frac{2\zeta}{5 \pi} (3 \pi^2 n)^{4/3}
\hspace{1cm} {\rm as\ } a \to \pm \infty,
\label{C-unitary}\\
&\longrightarrow& 4 \pi^2 n^2 a^2
\hspace{1.8cm} {\rm as\ } a \to 0^+,
\end{eqnarray}
\label{C-limits}
\end{subequations}
%-----------------
where $\zeta$ in Eq.~(\ref{C-unitary}) is a universal constant
whose value was estimated in Ref.~\cite{Tan0508}
to be approximately 1.

The most promising systems for testing the Tan relations
experimentally are cold trapped atoms near a Feshbach resonance.  
Some of the possibilities have been discussed by Tan 
in Refs.~\cite{Tan0505,Tan0508}.  The momentum distributions
$\rho_{\sigma} (\bm {k})$  can be measured by suddenly 
turning off the trapping potential and simultaneously
using the Feshbach resonance to change the scattering length 
to 0.  The cloud of atoms will expand and the initial momentum 
distribution can be determined from its density distributions 
after the expansion.  There should be a scaling region in which 
$\rho_{\sigma} (\bm {k})$ has the form in Eq.~(\ref{tail-Tan})
and the contact $C$ is simply the coefficient.  As pointed out 
by Tan in Ref.~\cite{Tan0505}, the energy relation in 
Eq.~(\ref{energy-Tan}) can be used to determine the sum 
$\langle T \rangle + \langle I \rangle$ of the kinetic and
interaction energies from the momentum 
distribution.  The potential energy $\langle V \rangle$ can be 
determined from the density distributions before the expansion.
The virial theorem in Eq.~(\ref{virial-Tan}) is a nontrivial 
relation between $\langle T \rangle + \langle I \rangle$,
$\langle V \rangle$, and $C$ that can be tested experimentally.
The adiabatic relation in Eq.~(\ref{adiabatic-Tan}) can also be 
tested by using the Feshbach resonance to change the scattering 
length and then measuring the changes in 
$\langle T \rangle + \langle I \rangle$,
$\langle V \rangle$, and $C$.

%%%%%%%%%%%%%%%%%%%%%%%%%%%%%%%%%%%%%%%%%%%%%%%%%%%%%%%%%%%%%
\section{Zero-Range Model}
\label{sec:ZRM}
%%%%%%%%%%%%%%%%%%%%%%%%%%%%%%%%%%%%%%%%%%%%%%%%%%%%%%%%%%%%%%%%%%%%%%%%%%

In Ref.~\cite{Braaten:2008uh}, the Tan relations were rederived 
using the quantum field theory for the Zero-Range Model.
They follow straightforwardly from standard renormalization 
methods and from the operator product expansion (OPE).
This analysis revealed that the contact $C$ can be expressed
as the integral over space of the expectation value of a local 
operator. In this section, we summarize the derivations in 
Ref.~\cite{Braaten:2008uh}.

A quantum field theory that describe atoms 
with two spin states must have
two quantum fields $\psi_\sigma(\bm{r})$, $\sigma = 1, 2$.
The Hamiltonian for a local quantum field theory can be expressed 
as the integral over
space of a Hamiltonian density: $H =\int d^3r \, {\cal H}$.
If the atoms are in an external potential $V(\bm{r})$,
the Hamiltonian density is the sum of
a kinetic term ${\cal T}$, an interaction term ${\cal I}$, 
and an external potential term ${\cal V}$:
${\cal H} = {\cal T} + {\cal I} + {\cal V}$.
The simplest quantum field theory that describes atoms 
with a large scattering length is the {\it Zero-Range Model},
in which the phase shift has the simple form given in 
Eq.~(\ref{delta0:ZRM}) up to arbitrarily large momentum.
For the Zero-Range Model, the three terms 
in the Hamiltonian density are
%-----------------
\begin{subequations}
\begin{eqnarray}
{\cal T} &=& 
\sum_\sigma \frac{1}{2m} 
        \nabla \psi_\sigma^\dagger \cdot \nabla \psi_\sigma^{(\Lambda)},
\label{T}
\\
{\cal I} &=& 
\frac{\lambda_0(\Lambda)}{m} \psi_1^\dagger \psi_2^\dagger \psi_1 \psi_2^{(\Lambda)},
\label{I}
\\
{\cal V} &=& V(\bm{R}) \sum_\sigma \psi_\sigma^\dagger \psi_\sigma.
\label{V}
\end{eqnarray}
\label{H}
\end{subequations}
%-----------------
For simplicity, we have set $\hbar = 1$.
The superscripts $(\Lambda)$ on the operators in 
Eqs.~(\ref{T}) and (\ref{I})
indicate that their matrix elements are ultraviolet divergent 
and an ultraviolet cutoff is required to make them well defined. 
For the ultraviolet cutoff, we impose an upper limit 
$|\bm{k}|<\Lambda$ on the momenta of virtual particles.
In the limit $\Lambda \to \infty$, 
the Hamiltonian density in Eq.~(\ref{H}) describes atoms with 
the phase shift given by Eq.~(\ref{delta0:ZRM}) if we take the 
coupling constant to be
%-----------------
\begin{equation}
\lambda_0(\Lambda) = \frac{4 \pi a}{1 - 2 a \Lambda/\pi} .
\label{g2}
\end{equation}
%-----------------
The connected Green's function 
for the scattering of a pair of atoms 
with spins 1 and 2 has a well-behaved limit as $\Lambda \to \infty$:
%-----------------
\begin{equation}
{\cal A}(E) = \frac{4 \pi/m}{-1/a + \sqrt{- m E - i \epsilon}} ,
\label{A-E}
\end{equation}
%-----------------
where $E$ is the total energy of the two atoms 
in their center-of-mass frame.
The T-matrix element for scattering of a
pair of atoms with momenta $+\bm{p}$ and $-\bm{p}$
is obtained by setting $E = p^2/m$.
Comparing with the expression for the T-matrix element in 
Eq.~(\ref{T-k}), we obtain the phase shift in Eq.~(\ref{delta0:ZRM}).

In Ref.~\cite{Braaten:2008uh}, the contact $C$ in the Tan relations
was identified as the integral over space of the expectation value 
of the local operator 
$\lambda_0^2(\Lambda) 
\psi_1^\dagger \psi_2^\dagger \psi_1 \psi_2^{(\Lambda)}$.
Matrix elements of the operator, which is the product 
of four quantum fields, are ultraviolet divergent, 
as indicated by the superscript $(\Lambda)$, 
but the dependence of the matrix elements on $\Lambda$ 
is precisely cancelled by the prefactor $\lambda_0^2(\Lambda)$.
Since this operator has ultraviolet finite matrix elements, 
we suppress the dependence on $\Lambda$ and denote it by
$\lambda_0^2 \psi_1^\dagger \psi_2^\dagger \psi_1 \psi_2$.
The contact can be expressed as%
\footnote{
We denote the product of local
operators ${\cal O}_A (\bm{R})$ and ${\cal O}_B (\bm{R})$ at the same
point in space by ${\cal O}_A {\cal O}_B (\bm{R})$.}
%-----------------
\begin{equation}
C = \int \! d^3R \, 
\langle X | \lambda_0^2 \psi_1^\dagger \psi_2^\dagger 
          \psi_1 \psi_2(\bm{R}) | X \rangle  .
\label{contact-ZRM}
\end{equation}
%-----------------
This expression makes explicit the dependence of the contact 
on the state $| X \rangle$.  It also reveals that the contact 
is an extensive quantity.

For some of the Tan relations, the derivations in 
Ref.~\cite{Braaten:2008uh} followed in a straightforward way from
the renormalization of the quantum field theory
once one realizes that 
$\lambda_0^2 \psi_1^\dagger \psi_2^\dagger \psi_1 \psi_2$
is a finite operator.
For example, the {\it energy relation} in Eq.~(\ref{energy-Tan}) 
can be derived by 
separating the interaction term 
$\int \! d^3 R \, \langle X | {\cal I} | X \rangle$ into two terms 
by using the identity
%-----------------
\begin{equation}
\lambda_0 =
- \frac{\Lambda}{2 \pi^2} \lambda_0^2
+ \frac{1}{4 \pi a} \lambda_0^2 .
\label{lambda0}
\end{equation}
%-----------------
The {\it adiabatic relation} in Eq.~(\ref{adiabatic-Tan}) 
can be derived from the Feynman-Hellman theorem,
%-----------------
\begin{equation}
d E/da = 
\int \!\! d^3R \, \langle X | \partial{\cal H}/\partial a | X \rangle,
\label{E-a}
\end{equation}
%-----------------
together with the identity
%-----------------
\begin{equation}
\frac{d \ }{da} \lambda_0 =
\frac{1}{4 \pi a^2} \lambda_0^2 .
\label{dlambda0}
\end{equation}
%-----------------
The {\it pressure relation} in Eq.~(\ref{pressure-Tan}) can be derived 
by applying dimensional analysis to the free energy density
(or thermodynamic potential density)
${\cal F}$, which requires
%-----------------
\begin{equation}
\left[ T \frac{\partial \ }{\partial T} 
- \frac12 a \frac{\partial \ }{\partial a} 
\right] {\cal F} = \frac52 {\cal F} .
\label{F-diman}
\end{equation}
%-----------------
The {\it virial theorem} in Eq.~(\ref{virial-Tan}) can be derived 
by applying dimensional analysis to the energy
$E = \int d^3R \, \langle X | {\cal H} | X \rangle$, which requires
%-----------------
\begin{equation}
\left[ \omega \frac{\partial \ }{\partial \omega} 
- \frac12 a \frac{\partial \ }{\partial a} 
\right] E = E .
\label{E-diman}
\end{equation}
%-----------------
The expression for the {\it inelastic loss rate} $\Gamma$
in Eq.~(\ref{Gam-Tan}) can be derived from the adiabatic relation 
in Eq.~(\ref{adiabatic-Tan}) by identifying $- \frac12 \Gamma$ as the 
imaginary part of the energy $E$ that results from adding a 
small negative imaginary part to the scattering length $a$ 
to take into account the effects of the inelastic scattering
channels.

The derivations of the tail of the momentum distribution 
in Eq.~(\ref{tail-Tan}) and the expression for the 
contact density ${\cal C}$ in terms 
of the local pair density in Eq.~(\ref{pairdensity-Tan})
require the OPE.
The number density operator for atoms with spin $\sigma$ 
is $\psi_\sigma^\dagger \psi_\sigma(\bm{R})$.
The momentum distribution for atoms with spin $\sigma$ 
in the state $| X \rangle$ can be expressed as 
%-----------------
\begin{equation}
\rho_{\sigma} (\bm{k}) = 
\langle X | \tilde \psi_\sigma^\dagger(\bm{k}) 
           \tilde \psi_\sigma(\bm{k}) | X \rangle ,
\label{rho-psitilde}
\end{equation}
%-----------------
where $\tilde \psi_\sigma(\bm{k})$ is the Fourier transform 
of the quantum field $\psi_\sigma(\bm{r})$.  The expression 
for the momentum distribution in terms of the quantum field is
%-----------------
\begin{equation}
\rho_{\sigma} (\bm{k}) = 
\int \! d^3 R \, \int \! d^3r \,  e^{i \bm{k} \cdot \bm{r}} 
\langle X | \psi_\sigma^\dagger(\bm{R} \mbox{$-\frac12$} \bm{r})
\psi_\sigma(\bm{R} \mbox{$+\frac12$} \bm{r}) | X \rangle .
\label{rho-psi}
\end{equation}
%-----------------
The OPE for the bilocal operator inside 
the expectation value has the form
%-----------------
\begin{equation}
\psi_\sigma^\dagger(\bm{R} - \mbox{$\frac12$} \bm{r}) 
\psi_\sigma(\bm{R} + \mbox{$\frac12$} \bm{r}) = 
\sum_n C_{\sigma,n}(\bm{r}) {\cal O}_n (\bm{R}) ,
\label{OPE}
\end{equation}
%-----------------
where the sum is over all possible local operators.
The local operators can be expressed as products of any number 
of quantum fields $\psi_\sigma(\bm{R})$ and an equal number of
quantum fields $\psi_\sigma^\dagger(\bm{R})$, 
with any number of gradients applied to those fields.
If the OPE in Eq.~(\ref{OPE}) is inserted into Eq.~(\ref{rho-psi}),
the momentum distribution reduces to
%-----------------
\begin{equation}
\rho_{\sigma} (\bm{k}) = 
\sum_n \left( \int \! d^3r \,  e^{i \bm{k} \cdot \bm{r}} 
	C_{\sigma,n}(\bm{r}) \right)
\int \! d^3 R \, \langle X | {\cal O}_n (\bm{R}) | X \rangle .
\label{rho-ope}
\end{equation}
%-----------------
Since the OPE is an asymptotic expansion for small $\bm{r}$,
Eq.~(\ref{rho-ope}) is an asymptotic expansion for large $\bm{k}$.
If a Wilson coefficient $C_{\sigma,n}(\bm{r})$ can be expanded 
as a power series in the vector $\bm{r}$, the corresponding 
integral in Eq.~(\ref{rho-ope}) can be expressed in terms of 
the Dirac delta function in $\bm{k}$ and derivatives of 
the Dirac delta function.  A power-law tail in
$\rho_{\sigma} (\bm{k})$ at large $\bm{k}$ can arise only 
from a term whose Wilson coefficient is not an analytic
function of the vector $\bm{r}$ at  $\bm{r} = 0$.

As shown in Ref.~\cite{Braaten:2008uh},
the OPE in Eq.~(\ref{OPE}) can be expressed as
%-----------------
\begin{eqnarray}
\psi_\sigma^\dagger(\bm{R} - \mbox{$\frac12$} \bm{r}) 
\psi_\sigma(\bm{R} + \mbox{$\frac12$} \bm{r}) &=& 
\psi_\sigma^\dagger \psi_\sigma(\bm{R})
+ \mbox{$\frac12$} \bm{r} \cdot
\left[ \psi_\sigma^\dagger \nabla \psi_\sigma(\bm{R}) 
- \nabla \psi_\sigma^\dagger \psi_\sigma(\bm{R}) \right]
\nonumber
\\
&& \hspace{2cm}
- \frac{r}{8 \pi} 
\lambda_0^2 \psi_1^\dagger \psi_2^\dagger \psi_1 \psi_2(\bm{R})
+ \ldots  ,
\label{OPE-ZRM}
\end{eqnarray}
%-----------------
where we have written explicitly all terms whose Wilson 
coefficients go to zero more slowly than $r^2$ as $r \to 0$.
The first two terms on the right side of the OPE in 
Eq.~(\ref{OPE-ZRM})
can be obtained by multiplying the Taylor 
expansions of the two operators.
The third term arises from quantum fluctuations 
involving pairs of atoms with small separations.
Its Wilson coefficient is proportional 
to $r = |\bm{r}|$, which is not an analytic
function of the vector $\bm{r}$ at $\bm{r} = 0$. 
Its Fourier transform at nonzero values of $\bm{k}$ is given by
%-----------------
\begin{equation}
\int \! d^3r \,  r e^{i \bm{k} \cdot \bm{r}} =
- \frac{8 \pi}{k^4} .
\label{r-ft}
\end{equation}
%-----------------
This can be derived by differentiating the Fourier transform of $1/r$.
The corresponding term in Eq.~(\ref{rho-ope})
gives a power-law tail in the momentum distribution:
%-----------------
\begin{equation}
\rho_{\sigma} (\bm{k}) \longrightarrow 
\int \! d^3 R \, \langle X | \lambda_0^2 
\psi_1^\dagger \psi_2^\dagger \psi_1 \psi_2 (\bm{R}) | X \rangle/k^4 .
\label{rho-k4}
\end{equation}
%-----------------
Comparing with Eq.~(\ref{tail-Tan}), we obtain the expression 
for the contact $C$ in Eq.~(\ref{contact-ZRM}).

The expression for $C$ in terms of the local pair density 
in Eq.~(\ref{pairdensity-Tan}) follows from the OPE for 
$\psi_1^\dagger \psi_1(\bm{R} + \bm{r})$ and 
$\psi_2^\dagger \psi_2(\bm{R} + \bm{r}')$.
In Ref.~\cite{Braaten:2008uh}, it was shown that the OPE 
includes the operator 
$\lambda_0^2 \psi_1^\dagger \psi_2^\dagger \psi_1 \psi_2$ 
with a Wilson coefficient 
proportional to $|\bm{r}' - \bm{r}|^{-2}$:
%-----------------
\begin{equation}
\psi_1^\dagger \psi_1(\bm{R} + \bm{r}) 
\psi_2^\dagger \psi_2(\bm{R} + \bm{r}') = 
\frac{1}{16 \pi^2 |\bm{r}' - \bm{r}|^2} \lambda_0^2 
\psi_1^\dagger \psi_2^\dagger \psi_1 \psi_2 (\bm{R}) 
+ \ldots  .
\label{OPE-nn}
\end{equation}
%-----------------
Only the most singular term as $\bm{r}' \to \bm{r}$ 
is shown explicitly on the right side.  Integrating the left side 
over $\bm{r}'$ and $\bm{r}$ inside a small ball of radius $s$,
we obtain an operator that counts the number of pairs of atoms 
with spins 1 and 2 inside that ball.  
If the right side of Eq.~(\ref{OPE-nn}) 
is integrated over the same region and
multiplied by $4 \pi/s$, the only term that survives 
in the limit $s \to 0$ is the one that is shown explicitly. 
Taking the expectation value of both sides, 
we obtain the expression for the contact density $\cal{C}$
in Eq.~(\ref{pairdensity-Tan}).

%%%%%%%%%%%%%%%%%%%%%%%%%%%%%%%%%%%%%%%%%%%%%%%%%%%%%%%%%%%%%
\section{Resonance Model}
\label{sec:RM}
%%%%%%%%%%%%%%%%%%%%%%%%%%%%%%%%%%%%%%%%%%%%%%%%%%%%%%%%%%%%%%%%%%%%%%%%%%

A practical method for controlling the scattering length of atoms 
is by exploiting Feshbach resonances.  A Feshbach resonance 
occurs at a value $B_0$ of the magnetic field for which a 
molecular state in a closed channel with a higher two-atom threshold
is in resonance with a 
pair of atoms at threshold in the channel of interest.
Near the Feshbach resonance, the dependence of the 
scattering length $a$ on the magnetic field $B$ 
can be approximated by
\begin{eqnarray}
a(B) = a_{\rm bg} \left( 1 - \frac {\Delta} {B-B_0} \right) ,
\label{a:FR}
\end{eqnarray}
where $a_{\rm bg}$ is the background scattering length far 
from the resonance and $B_0 +\Delta$ is the position 
of a zero of the scattering length.

A natural description for atoms near a Feshbach resonance 
is provided by the {\it Resonance Model}, in which the molecule 
responsible for the resonance is treated as a point particle.
The Resonance Model has three parameters $\nu$, $g$, and $\lambda$ 
that can be defined by the S-wave phase shift 
given in Eq.~(\ref{delta0:RM}).
The scattering length and the effective range are
\begin{subequations}
\begin{eqnarray}
a &=&  \frac{1}{4 \pi} \left( \lambda - \frac{g^2}{m \nu} \right) ,
\label{a:RM}
\\
r_s &=&  - 8 \pi \left( \lambda - \frac{g^2}{m\nu} \right)^{-2}
\frac{g^2}{m^2 \nu^2}.
\label{r:RM}
\end{eqnarray}
\end{subequations}
The effective range is negative definite.
The scattering length in Eq.~(\ref{a:RM}) can be made 
arbitrarily large by tuning $\nu$ to near 0.  
In this case, the limiting value of the effective range is 
$r_s \to -8 \pi/g^2$.
The standard expression for the scattering length near a Feshbach 
resonance in Eq.~(\ref{a:FR}) can be reproduced 
by taking $\nu$ to be linear in the magnetic field $B$ 
while $g^2$ and $\lambda$ are constants:
\begin{subequations}
\begin{eqnarray}
\nu  & = & -\mu \, (B - B_0),
\label{nu-phys:RM}
\\
g^2   & = & - 4 \pi m a_{\rm bg} \, \mu \, \Delta,
\\
\lambda & = & 4 \pi a_{\rm bg}.
\end{eqnarray}
\label{param-phys:RM}
\end{subequations}
The parameter $\mu$ in Eqs.~(\ref{param-phys:RM}) 
is the difference between the magnetic moment 
of the molecule and twice the magnetic moment of an isolated atom.
Away from the Feshbach resonance, the parameter $\nu$ 
can be interpreted as the detuning energy 
of the molecular state relative to the two-atom threshold.
Near the Feshbach resonance, the strong coupling 
between the molecular state and two-atom scattering states 
makes the interpretation of $\nu$ more complicated.

%%%%%%%%%%%%%%%%%%%%%%%%%%%%%%%%%%%%%%%%%%%%%%%%%%
\begin{figure}[t]
\centerline{\includegraphics*[height=5cm,angle=0,clip=true]{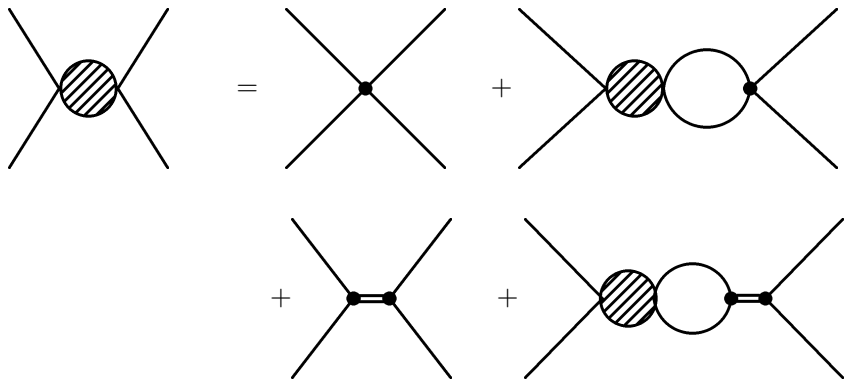}}
\vspace*{0.0cm}
\caption{Integral equation for the scattering of a 
pair of atoms in the Resonance Model.   Single lines denote
atom propagators and double lines denote molecule propagators.
The blob represents the amplitude $i{\cal A}(E)$.
The solid dots represent the interaction vertices 
$-i \lambda_0/m$ and $-i g_0/m$.
}
\label{fig:inteqRM}
\end{figure}
%%%%%%%%%%%%%%%%%%%%%%%%%%%%%%%%%%%%%%%%%%%%%%%%%%

The Resonance Model was first formulated as a local quantum field theory 
by Kaplan \cite{Kaplan:1996nv} and by Kokkelmans et al.~\cite{KMCWH02}.
In the sector that consists of states containing two atoms 
or one diatomic molecule, the model can be solved analytically.
The solution reveals that the model is renormalizable,
at least in this sector. 
Kaplan used dimensional regularization to remove ultraviolet
divergences, thus avoiding the need for explicit 
renormalization of the parameters \cite{Kaplan:1996nv}.
Kokkelmans et al.~\cite{KMCWH02} derived the
renormalizations of the parameters that are required to make 
the observables independent of the ultraviolet momentum cutoff.
They used numerical solutions of mean-field integral equations 
for this model to demonstrate the feasibility 
for achieving superfluidity in ultracold gas of fermionic atoms.

In the quantum field theory formulation of the Resonance Model
for fermionic atoms with two spin states,
there are three quantum fields: fermionic fields $\psi_1$ and $\psi_2$
that annihilate atoms
and a bosonic field $\phi$ that annihilates a diatomic molecule.
The Hamiltonian density for the Resonance Model  is the sum of
a kinetic term ${\cal T}$, an interaction term ${\cal I}$, 
and an external potential term ${\cal V}$:
${\cal H} = {\cal T} + {\cal I} + {\cal V}$.
The individual terms are
%-----------------
\begin{subequations}
\begin{eqnarray}
{\cal T} & = &
\frac{1}{2m} \nabla \psi_1^\dagger \cdot \nabla \psi_1^{(\Lambda)}
+ \frac{1}{2m} \nabla \psi_2^\dagger \cdot \nabla \psi_2^{(\Lambda)}
+ \frac{1}{4m} \nabla \phi^\dagger \cdot \nabla \phi ,
\label{T:RM}
\\
{\cal I} & = &  \nu_0 \phi^\dagger \phi
+ \frac{g_0}{m} \left(\phi^\dagger \psi_1 \psi_2^{(\Lambda)} 
+ \psi_1^\dagger \psi_2^\dagger \phi^{(\Lambda)} \right)
+ \frac{\lambda_0}{m} \psi_1^\dagger \psi_2^\dagger \psi_1 \psi_2^{(\Lambda)} ,
\label{I:RM}
\\
{\cal V} &=& V(\bm{R}) 
\left( \psi_1^\dagger \psi_1 + \psi_2^\dagger \psi_2
+ 2 \phi^\dagger \phi \right) .
\label{V:RM}
\end{eqnarray}
\label{TIV:RM}
\end{subequations}
%-----------------
To avoid ultraviolet divergences, an ultraviolet cutoff $\Lambda$
must be imposed on the momenta of virtual particles.
The superscripts $(\Lambda)$ on some of the operators in 
Eqs.~(\ref{T:RM}) and (\ref{I:RM})
indicate that they have matrix elements that are ultraviolet divergent. 
For the model to have a nontrivial finite limit as $\Lambda \to \infty$,
the parameters $\nu_0$, $g_0$, and $\lambda_0$ 
in the interaction term in Eq.~(\ref{I:RM}) must depend on $\Lambda$.
We will refer to them as the {\it bare parameters}.
The parameters $\nu$, $g$, and $\lambda$ in the phase shift 
in Eq.~(\ref{delta0:RM}) do not depend on $\Lambda$.
We will refer to them as the {\it renormalized parameters}.
The values for the bare parameters that are required to
reproduce the phase shift in Eq.~(\ref{delta0:RM}) are
\begin{subequations}
\begin{eqnarray}
\lambda_0 (\Lambda) & = & Z(\Lambda) \lambda,
\label{lambda0:RM}
\\
g_0 (\Lambda) & = & Z(\Lambda) g ,
\label{g0:RM}
\\
\nu_0 (\Lambda) & = & \nu - [1-Z(\Lambda)] \frac{g^2}{m\lambda},
\label{nu0:RM}
\end{eqnarray}
\label{param0:RM}
\end{subequations}
where the renormalization constant $Z$ is
\begin{eqnarray}
Z(\Lambda) = \left( 1 - \frac{\lambda \Lambda}{2 \pi^2} \right)^{-1}.
\label{Z-Lambda}
\end{eqnarray}
There are two simple combinations of the parameters 
that are renormalization invariants:
\begin{subequations}
\begin{eqnarray}
\frac{g}{\lambda} & = & \frac{g_0}{\lambda_0},
\label{g-lam:rgi}
\\
\nu - \frac{g^2}{m \lambda}  & = & \nu_0 - \frac{g_0^2}{m \lambda_0}.
\label{nu-g-lam:rgi}
\end{eqnarray}
\label{rgi:RM}
\end{subequations}

The amplitude for the scattering of a pair of atoms can be calculated 
by solving the Lippmann-Schwinger integral equation, 
which is represented diagrammatically in Fig.~\ref{fig:inteqRM}. 
After evaluating the loop integral using Eq.~(\ref{intq1}),
the integral equation reduces to an algebraic equation:
%-----------------
\begin{equation}
{\cal A} (E) = - \frac{1}{m} 
\left ( \lambda_0 + \frac{g_0^2/m}{E- \nu_0 + i \epsilon} \right)
\left[ 1 + m {\cal A} (E) 
\left( \frac{\Lambda}{2 \pi^2} 
	- \frac{1}{4 \pi} \sqrt{- m E - i \epsilon} \right) \right] ,
\label{A-E:inteq}
\end{equation}
%-----------------
where $\Lambda$ is an ultraviolet momentum cutoff.
The solution ${\cal A}(E)$ depends on the total energy $E$ of the 
pair of atoms in the center-of-mass frame and not 
separately on their momenta. 
After substituting Eqs.~(\ref{param0:RM}) for the bare parameters, 
the amplitude becomes independent of the ultraviolet cutoff:
%-----------------
\begin{equation}
{\cal A} (E) = - \frac{1}{m} \left[
\left ( \lambda + \frac{g^2/m}{E- \nu + i \epsilon} \right)^{-1}
- \frac{1}{4 \pi} \sqrt{- m E - i \epsilon} \right ]^{-1}  .
\label{A-E:RM}
\end{equation}
%-----------------
The T-matrix element for scattering of a
pair of atoms with momenta $+\bm{p}$ and $-\bm{p}$
is obtained by setting $E = p^2/m$.
Comparing with the expression for the T-matrix element in 
Eq.~(\ref{T-k}), we obtain the phase shift in Eq.~(\ref{delta0:RM}).

Since the Resonance Model can be formulated as a quantum 
field theory, the methods used to derive the Tan relations 
in Ref.~\cite{Braaten:2008uh} can be applied equally well to this model.
Because the Resonance Model is a local quantum field theory,
we can use the OPE to identify the
contact density operator.
Because it is renormalizable,
we can use standard renormalization methods to derive 
the analogs of some of the other Tan relations.
In the following seven sections,
we deduce the generalized Tan relations for the  
Resonance Model.

%%%%%%%%%%%%%%%%%%%%%%%%%%%%%%%%%%%%%%%%%%%%%%%%%%%%%%%%%%%%%%%%%%%%%%%%%%
\section{Tail of the momentum distribution}
\label{sec:tail}
%%%%%%%%%%%%%%%%%%%%%%%%%%%%%%%%%%%%%%%%%%%%%%%%%%%%%%%%%%%%%%%%%%%%%%%%%%

The Tan relation in Eq.~(\ref{tail-Tan}) states that the 
momentum distributions $\rho_{\sigma} (\bm {k})$ 
have power-law tails proportional to $1/k^4$, 
with the coefficient for both spin states given by the contact $C$.  
We will use the operator product expansion (OPE)
to show that the momentum distributions 
in the Resonance Model also have power-law tails proportional 
to $1/k^4$ and we will express the contact $C$ as the 
integral over space of the expectation value of a local operator.

Our starting point is the expression for the momentum distribution 
$\rho_{\sigma} (\bm {k})$ in Eq.~(\ref{rho-psi}), 
which involves the expectation value of
$\psi_\sigma^\dagger(\bm{R} - \frac12 \bm{r})
 \psi_\sigma(\bm{R} + \frac12 \bm{r})$.
Inserting the OPE in Eq.~(\ref{OPE}) for this bilocal operator,
we obtain the asymptotic expansion of $\rho_{\sigma} (\bm {k})$
for large $\bm{k}$ in Eq.~(\ref{rho-ope}).  A power-law tail in
$\rho_{\sigma} (\bm {k})$ can only come from terms in the OPE 
whose Wilson coefficients $C_{\sigma,n}(\bm{r})$ are not 
analytic functions of $\bm{r}$ at $\bm{r}=0$.
Such terms can arise from short-distance quantum fluctuations 
involving pairs of atoms.

The Wilson coefficients on the right side of 
the general OPE in Eq.~(\ref{OPE-gen})
can be determined by a procedure called {\it matching}.
First, matrix elements of the bilocal operator 
${\cal O}_A(\bm{R} - \frac12 \bm{r}) 
 {\cal O}_B(\bm{R} + \frac12 \bm{r})$ between appropriate 
states are calculated and expanded in powers of $r = |\bm{r}|$.
Second, matrix elements of the local operators 
${\cal O}_C (\bm{R})$ between the same states are calculated.
Finally, the coefficients $C_{A,B}^C(\bm{r})$ are determined by 
demanding that the expansions in powers of $r$ on 
both sides of the equation match.  The coefficient of
${\cal O}_C (\bm{R})$ can be determined by matching matrix elements 
between any states $|X \rangle$ and $|X' \rangle$ for which
$\langle X' | {\cal O}_C (\bm{R}) | X \rangle$ is nonzero.
We can exploit the fact that the OPE is an operator equation
by choosing the simplest possible states for the matching.

%%%%%%%%%%%%%%%%%%%%%%%%%%%%%%%%%%%%%%%%%%%%%%%%%%
\begin{figure}[t]
\centerline{\includegraphics*[height=2.5cm,angle=0,clip=true]{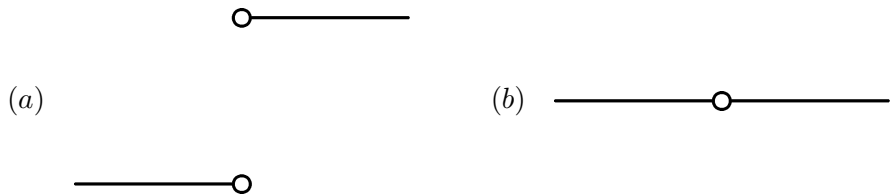}}
\vspace*{0.0cm}
\caption{Diagrammatic representations of
(a) the bilocal operator
$\psi_\sigma^\dagger (\bm{R} - \mbox{$\frac12$} \bm{r}) 
	\psi_\sigma(\bm{R} + \mbox{$\frac12$} \bm{r})$
and (b) local operators that annihilate one atom and create one atom,
such as $\psi_\sigma^\dagger \psi_\sigma(\bm{R})$.
}
\label{fig:vertop}
\end{figure}
%%%%%%%%%%%%%%%%%%%%%%%%%%%%%%%%%%%%%%%%%%%%%%%%%%

%%%%%%%%%%%%%%%%%%%%%%%%%%%%%%%%%%%%%%%%%%%%%%%%%%
\begin{figure}[t]
\centerline{\includegraphics*[height=5cm,angle=0,clip=true]{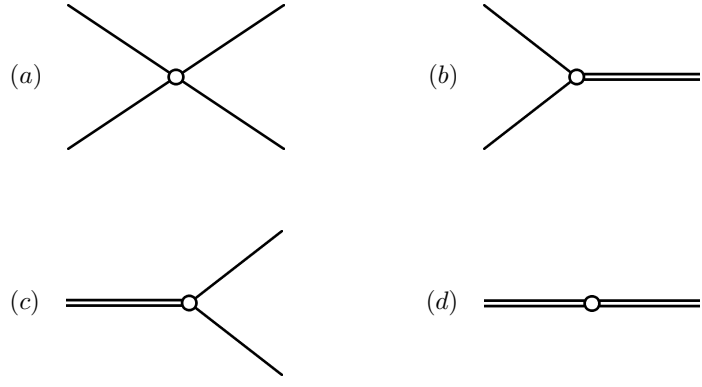}}
\vspace*{0.0cm}
\caption{Diagrammatic representations of the local operators 
(a) $\psi_1^\dagger \psi_2^\dagger \psi_1 \psi_2$,
(b) $\psi_1^\dagger \psi_2^\dagger \phi$, 
(c) $\phi^\dagger \psi_1 \psi_2$, and 
(d) $\phi^\dagger \phi$.
}
\label{fig:linepoint}
\end{figure}
%%%%%%%%%%%%%%%%%%%%%%%%%%%%%%%%%%%%%%%%%%%%%%%%%%

The matrix elements of the operators in the OPE 
can be evaluated using diagrammatic methods by applying
the Feynman rules given in Appendix~\ref{app:diagrams}.
The bilocal operator 
$\psi_\sigma^\dagger(\bm{R} - \frac12 \bm{r})
 \psi_\sigma(\bm{R} + \frac12 \bm{r})$
can be represented diagrammatically by a pair of open dots
with an atom line ending at the dot associated 
with $\psi_\sigma$ and an atom line
beginning at the dot associated with 
$\psi_\sigma^\dagger$, as in Fig.~\ref{fig:vertop}(a).
The atom lines should be labelled by the spin $\sigma$, 
but we suppress that label.
Local operators that annihilate one atom and 
create one atom, such as
$\psi_\sigma^\dagger \psi_\sigma$ and 
$\psi_\sigma^\dagger \nabla^j \psi_\sigma
	- \nabla^j \psi_\sigma^\dagger \psi_\sigma$,
can be represented diagrammatically by a single open dot, 
with an atom line ending at the dot and another atom line 
beginning at the dot, as in Fig.~\ref{fig:vertop}(b).   
Local operators that annihilate
a pair of atoms with spins 1 and 2 or a molecule
and create a pair of atoms or a molecule, such as
$\psi_1^\dagger \psi_2^\dagger \psi_1 \psi_2$,
$\psi_1^\dagger \psi_2^\dagger \phi$, $\phi^\dagger \psi_1 \psi_2$,
and $\phi^\dagger \phi$,
can be represented diagrammatically by a single open dot, 
with two atom lines or a molecule line 
ending at the dot 
and two atom lines or a molecule line 
beginning at the dot, as in Fig.~\ref{fig:linepoint}.

The Wilson coefficients for operators that annihilate 
one atom and create one atom
can be obtained by matching matrix elements between
the 1-atom states 
consisting of a single atom with specified momentum and spin.
For the bilocal operator 
$\psi_\sigma^\dagger(\bm{R} - \frac12 \bm{r})
 \psi_\sigma(\bm{R} + \frac12 \bm{r})$,
the matrix element between 1-atom states 
is given by the diagram in Fig.~\ref{fig:vertop}(a).
The expression for the matrix element is simply the 
Feynman rule for the operator, which is an exponential 
function of $\bm{r}$.
For local operators that annihilate one atom and 
create one atom, such as
$\psi_\sigma^\dagger \psi_\sigma(\bm{R})$,
the matrix element is given by the diagram in 
Fig.~\ref{fig:vertop}(b).
Their Wilson coefficients can be determined by matching 
the matrix elements between the 1-atom states.
The Wilson coefficients are identical to 
those that would be obtained by multiplying the Taylor expansions 
of the operators
$\psi_\sigma^\dagger(\bm{R} - \frac12 \bm{r})$ and
$\psi_\sigma(\bm{R} + \frac12 \bm{r})$.

%%%%%%%%%%%%%%%%%%%%%%%%%%%%%%%%%%%%%%%%%%%%%%%%%%
\begin{figure}[t]
\centerline{\includegraphics*[height=6cm,angle=0,clip=true]{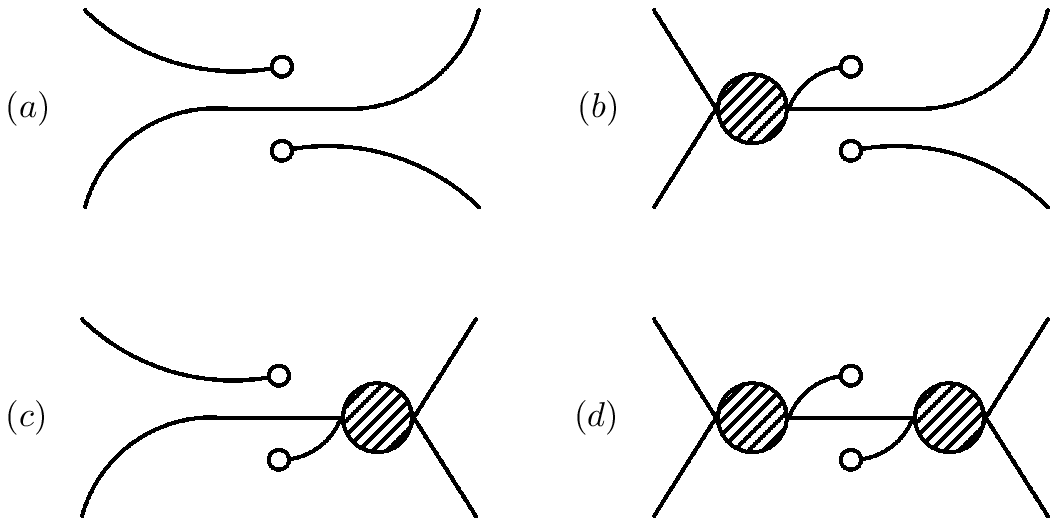}}
\vspace*{0.0cm}
\caption{The diagrams for matrix elements of the 
bilocal operator
$\psi_\sigma^\dagger(\bm{R} - \mbox{$\frac12$} \bm{r}) 
\psi_\sigma(\bm{R} + \mbox{$\frac12$} \bm{r})$
between 2-atom scattering states.
The open dots represent the operators.  }
\label{fig:<psipsibi>}
\end{figure}
%%%%%%%%%%%%%%%%%%%%%%%%%%%%%%%%%%%%%%%%%%%%%%%%%%

\begin{figure}[t]
\centerline{\includegraphics*[height=6cm,angle=0,clip=true]{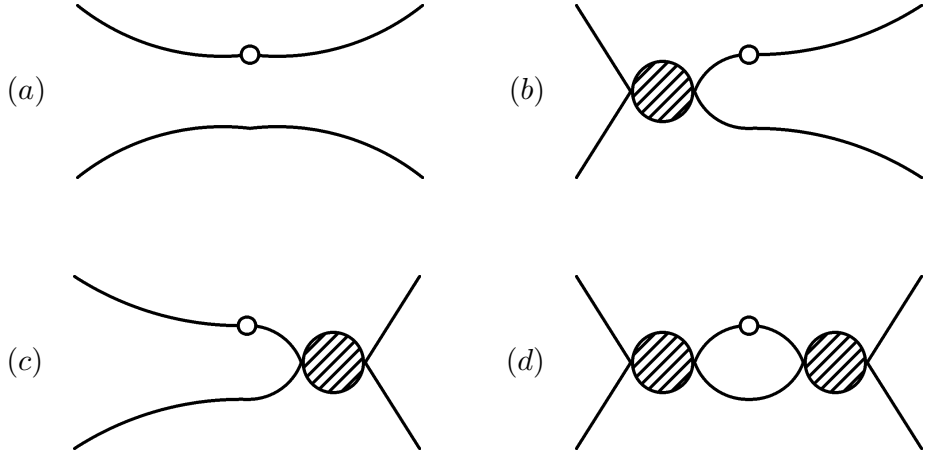}}
\vspace*{0.0cm}
\caption{The diagrams for matrix elements of local operators
that annihilate an atom and create an atom, 
such as $\psi_\sigma^\dagger \psi_\sigma$
or $\psi_\sigma^\dagger \nabla \psi_\sigma
	- \nabla \psi_\sigma^\dagger \psi_\sigma$, 
between 2-atom scattering states.
The open dot represents the operator.  }
\label{fig:<psipsi>}
\end{figure}

The Wilson coefficients of local operators that annihilate 
one atom and create one atom, such as 
$\psi_\sigma^\dagger \psi_\sigma(\bm{R})$, could also be determined 
by matching matrix elements between 2-atom states,
but the calculations are more complicated.
The diagrams for the matrix element of the bilocal operator
$\psi_\sigma^\dagger(\bm{R} - \frac12 \bm{r})
	\psi_\sigma(\bm{R} + \frac12 \bm{r})$
are shown in Fig.~\ref{fig:<psipsibi>}.  The blobs represent the
infinite series of diagrams that are summed up by the
integral equation in Fig.~\ref{fig:inteqRM}. 
The contributions from each of the three diagrams in 
Fig.~\ref{fig:<psipsibi>}(a,b,c) are analytic functions of $\bm{r}$.
The diagram in Fig.~\ref{fig:<psipsibi>}(d)
is evaluated in subsection~\ref{sec:me-psipsi-bilocal} 
of the Appendix.  It is an exponential function of $r = |\bm{r}|$, 
so it is not analytic at $\bm{r} = 0$.
The diagrams for the matrix element of a local operator
that annihilates one atom and creates one atom, 
such as $\psi_\sigma^\dagger \psi_\sigma(\bm{R})$,
are shown in Fig.~\ref{fig:<psipsi>}.  
Given the Wilson coefficients determined by matching matrix elements
between 1-atom states, the three diagrams in 
Fig.~\ref{fig:<psipsi>}(a,b,c) match the expansions in powers of
$\bm{r}$ of the corresponding diagrams in 
Fig.~\ref{fig:<psipsibi>}(a,b,c).  The diagram in 
Fig.~\ref{fig:<psipsi>}(d) matches the even powers of $r$ in the 
expansion of the diagram in 
Fig.~\ref{fig:<psipsibi>}(d) in powers of $r$.
The diagram in Fig.~\ref{fig:<psipsibi>}(d)
is evaluated in subsection~\ref{sec:me-psipsi-local} of the Appendix.
However for operators whose Wilson coefficient is an odd 
function of $\bm{r}$, such as 
$\psi_\sigma^\dagger \nabla \psi_\sigma
- \nabla \psi_\sigma^\dagger \psi_\sigma$,
the diagram in Fig.~\ref{fig:<psipsibi>}(d) vanishes.
The odd powers of $r$ in the expansion of the diagram in 
Fig.~\ref{fig:<psipsibi>}(d),
which are not analytic at $\bm{r}=0$, must be matched by more
complicated operators whose matrix elements between 1-atom states are zero.
The next most complicated operators 
with nontrivial matrix elements between 2-atom scattering 
states have one factor of
$\psi_1 \psi_2$ or $\phi$ and one factor of 
$\psi_1^\dagger \psi_2^\dagger$ or $\phi^\dagger$.

%%%%%%%%%%%%%%%%%%%%%%%%%%%%%%%%%%%%%%%%%%%%%%%%%%
%%%%%%%%%%%%%%%%%%%%%%%%%%%%%%%%%%%%%%%%%%%%%%%%%%

%%%%%%%%%%%%%%%%%%%%%%%%%%%%%%%%%%%%%%%%%%%%%%%%%%
\begin{figure}[t]
\centerline{\includegraphics*[height=3cm,angle=0,clip=true]{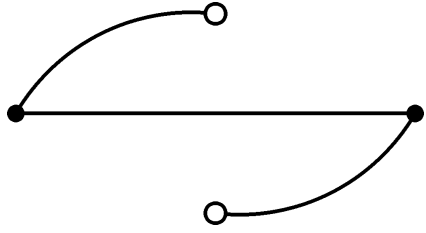}}
\vspace*{0.0cm}
\caption{The lines in the diagram of
Fig.~\ref{fig:<psipsibi>}(d) 
whose energies and momenta must be integrated over.
The solid dots are vertices to which are attached either 
a pair of atom lines or a molecule line.
}
\label{fig:lineint}
\end{figure}
%%%%%%%%%%%%%%%%%%%%%%%%%%%%%%%%%%%%%%%%%%%%%%%%%%

To identify the operators on the right side of the OPE 
whose matrix elements between 1-atom states are zero
but whose matrix elements between 2-atom scattering states 
are nonzero, we consider the lines in the diagram in 
Fig.~\ref{fig:<psipsibi>}(d) whose energies and momenta are integrated over.
Those lines are shown in Fig.~\ref{fig:lineint}, 
where the momenta are labelled.  
The solid dots are vertices whose Feynman rule
is $- i \lambda_0/m$ if a pair of atom lines is attached and 
$- i g_0/m$ if a molecule line is attached.  Nonanalytic 
dependence on $\bm{r}$ can arise only from the integration region 
with large momentum $\bm{q}$.  There can be a significant 
contribution from this region only if the coordinates associated 
with the endpoints of the lines carrying momentum $\bm{q}$ 
are close together.  Contributions from this region therefore 
correspond to shrinking those lines to a single point.  
If we include the lines attached to the solid dots in Fig.~\ref{fig:lineint},
the shrinking of the large-momentum lines to a point
gives one of the four vertices in Fig.~\ref{fig:linepoint}.  
These four vertices correspond to the operators 
$\psi_1^\dagger \psi_2^\dagger \psi_1 \psi_2$,
$\phi^\dagger \psi_1 \psi_2$, $\psi_1^\dagger \psi_2^\dagger \phi$,
and $\phi^\dagger \phi$, respectively.
They appear with relative weights $\lambda_0^2$, $\lambda_0 g_0$, 
$\lambda_0 g_0$, and $g_0^2$, respectively.
It is convenient to introduce a composite operator $\Phi(\bm{R})$
defined by
%-----------------
\begin{equation}
\Phi(\bm{R}) = \lambda_0 \psi_1 \psi_2^{(\Lambda)}(\bm{R}) + g_0 \phi(\bm{R}) .
\label{Phi-def}
\end{equation}
%-----------------
The simplest combination of operators that corresponds to the large-momentum 
region of the diagram in Fig.~\ref{fig:<psipsibi>}(d) is therefore 
$\Phi^\dagger \Phi(\bm{R})$. 

%%%%%%%%%%%%%%%%%%%%%%%%%%%%%%%%%%%%%%%%%%%%%%%%%%
\begin{figure}[t]
\centerline{\includegraphics*[height=6cm,angle=0,clip=true]{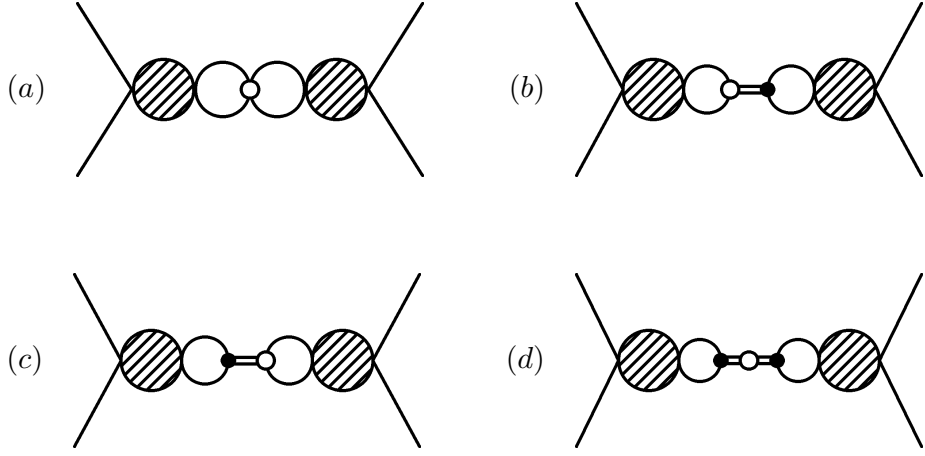}}
\vspace*{0.0cm}
\caption{Diagrams for matrix elements of the local operator
$\Phi^\dagger \Phi$ between 2-atom scattering states.
The open dot represents the operator. 
For each of the diagrams shown, there are three other diagrams 
with no scattering of the atoms in the initial state 
or in the final state or both.}
\label{fig:<PhiPhi>}
\end{figure}
%%%%%%%%%%%%%%%%%%%%%%%%%%%%%%%%%%%%%%%%%%%%%%%%%%

The Wilson coefficient of the $\Phi^\dagger \Phi(\bm{R})$ term 
can be determined by matching matrix elements
between two-atom scattering states.  The matrix element for 
$\Phi^\dagger \Phi(\bm{R})$ can be represented 
by the sum of the four diagrams in Fig.~\ref{fig:<PhiPhi>},
together with 12 other diagrams in which there is no scattering 
of the two incoming lines or no scattering of the two outgoing 
lines or both.  The sum of all these diagrams is
calculated in subsection~\ref{sec:me-PhiPhi} of the Appendix. 
The dependence on the initial
and final states is exactly what is required to match the 
term linear in $r$ in the expansion of the diagram in
Fig.~\ref{fig:<psipsibi>}(d).  By matching, we determine the
Wilson coefficient of the operator $\Phi^\dagger\Phi$ 
to be $- r/(8 \pi)$.  Thus the analog of the OPE in 
Eq.~(\ref{OPE-ZRM}) in the Resonance Model is
%-----------------
\begin{eqnarray}
\psi_\sigma^\dagger(\bm{R} - \mbox{$\frac12$} \bm{r}) 
\psi_\sigma(\bm{R} + \mbox{$\frac12$} \bm{r}) &=& 
\psi_\sigma^\dagger \psi_\sigma(\bm{R})
+ \mbox{$\frac12$} \bm{r} \cdot
\left[ \psi_\sigma^\dagger \nabla \psi_\sigma(\bm{R}) 
- \nabla \psi_\sigma^\dagger \psi_\sigma(\bm{R}) \right]
\nonumber
\\
&& \hspace{2cm}
- \frac{r}{8 \pi} \Phi^\dagger \Phi(\bm{R})
+ \ldots  ,
\label{OPE-RM}
\end{eqnarray}
%-----------------

We can now identify the coefficient $C$ of the 
$1/k^4$ tail in the momentum distribution in Eq.~(\ref{tail-Tan}). 
The leading term in the asymptotic expansion of the
momentum distribution in Eq.~(\ref{rho-ope}) comes from the term 
with operator $\Phi^\dagger\Phi$.  The Fourier integral of its 
Wilson coefficient $-r/(8 \pi)$ is given by Eq.~(\ref{r-ft}).
Comparing with Eq.~(\ref{tail-Tan}),
we find that the contact in the Resonance Model is
%-----------------
\begin{equation}
C = \int \! d^3R \, 
\langle X | \Phi^\dagger \Phi(\bm{R}) | X \rangle  ,
\label{contact-RM}
\end{equation}
%-----------------
where $\Phi(\bm{R})$ is the composite operator defined by
Eq.~(\ref{Phi-def}).  In subsequent sections, we will suppress the state 
$| X \rangle$ and denote the expectation value 
$\langle X | {\cal O} | X \rangle$ of an operator in that state 
simply by $\langle {\cal O} \rangle$.

%%%%%%%%%%%%%%%%%%%%%%%%%%%%%%%%%%%%%%%%%%%%%%%%%%%%%%%%%%%%%%%%%%%%%%%%%%
\section{Energy relation}
\label{sec:energy}
%%%%%%%%%%%%%%%%%%%%%%%%%%%%%%%%%%%%%%%%%%%%%%%%%%%%%%%%%%%%%%%%%%%%%%%%%%

Tan's energy relation in Eq.~(\ref{energy-Tan}) is a decomposition 
of the energy into terms that are separately insensitive 
to the range of the interactions between the atoms.
In a quantum field theoretic formulation of the problem,
the analogous relation is a decomposition of the energy 
into terms that are insensitive to the ultraviolet cutoff. 
To accomplish this separation for the Resonance Model,
we use the facts that $\phi^\dagger \phi$ 
and $\Phi^\dagger \Phi$ are operators with ultraviolet-finite 
matrix elements.  For $\Phi^\dagger \Phi$, this follows from 
it being the contact density operator.  For $\phi^\dagger \phi$, 
this follows from an adiabatic relation derived later in 
Section~\ref{sec:adiabatic}.  Using the expressions 
for the bare coupling constants in Eqs.~(\ref{param0:RM}),
the interaction term ${\cal I}$
in the Hamiltonian density in Eq.~(\ref{I:RM})
can be expressed in the form
%-----------------
\begin{equation}
{\cal I} = 
\left( \nu - \frac{g^2}{m \lambda} \right) \phi^\dagger \phi
+ \left( \frac{1}{m \lambda} - \frac{\Lambda}{2 \pi^2 m} \right)  
\Phi^\dagger \Phi .
\label{I-phiPhi}
\end{equation}
%-----------------
We can now decompose the sum of the kinetic energy density 
and the interaction energy density into five terms,
each of which has ultraviolet-finite matrix elements:
%-----------------
\begin{eqnarray}
{\cal T} + {\cal I} & = &
\frac{1}{2m} \sum_\sigma 
\left( \nabla \psi_\sigma^\dagger \cdot \nabla \psi_\sigma^{(\Lambda)}
      - \frac{\Lambda}{2 \pi^2} \Phi^\dagger \Phi \right)
+ \frac{1}{4m} \nabla \phi^\dagger \cdot \nabla \phi
\nonumber
\\
&&  + \left( \nu - \frac{g^2}{m \lambda} \right) \phi^\dagger \phi
+ \frac{1}{m \lambda} \Phi^\dagger \Phi .
\label{T+I-UVfinite}
\end{eqnarray}
%-----------------
To see that the term that is summed over $\sigma$  in
Eq.~(\ref{T+I-UVfinite}) has ultraviolet finite matrix elements, we 
consider the integral over space of its expectation value:
%-----------------
\begin{equation}
\langle T_\sigma^{\rm (sub)} \rangle = 
\frac{1}{2m} \int \! d^3R \, 
\left( \langle \nabla \psi_\sigma^\dagger 
	\cdot \nabla \psi_\sigma^{(\Lambda)} \rangle
      - \frac{\Lambda}{2 \pi^2} \langle \Phi^\dagger \Phi \rangle \right) .
\label{T-sub}
\end{equation}
%-----------------
This differs from the expectation value of the kinetic energy 
of atoms with spin $\sigma$ by a subtraction that is linear 
in the ultraviolet cutoff $\Lambda$.  By expressing the 
quantum fields $\psi$ and $\psi^\dagger$ in terms of their 
Fourier transforms and using the expression for the momentum 
distribution $\rho_\sigma(\bm{k})$ in Eq.~(\ref{rho-psitilde}),
we can express the subtracted kinetic energy 
in Eq.~(\ref{T-sub}) in the form 
%-----------------
\begin{equation}
\langle T_\sigma^{\rm (sub)} \rangle = 
\int \! \frac{d^3k}{(2 \pi)^3} \, \frac{k^2}{2m}
\left( \rho_\sigma(\bm{k}) - \frac{C}{k^4} \right) ,
\label{T-sub:k}
\end{equation}
%-----------------
where $C$ is the contact given in Eq.~(\ref{contact-RM}).
We have expressed the factor of
$\Lambda$ in Eq.~(\ref{T-sub}) as an integral over momentum space.
Since the tail of the momentum distributions at large $\bm{k}$ 
has the form in Eq.~(\ref{tail-Tan}), the subtraction in 
Eq.~(\ref{T-sub:k}) makes the integral over $\bm{k}$ convergent 
in the limit $\Lambda \to \infty$.  

The energy relation for the Resonance Model is obtained 
by taking the expectation value 
of ${\cal H} = {\cal T} + {\cal I} + {\cal V}$
and integrating over space.
It can be expressed in the form
%-----------------
\begin{eqnarray}
E &=& \langle V \rangle 
+ \langle T_1^{\rm (sub)} \rangle 
+ \langle T_2^{\rm (sub)} \rangle 
+ \langle T_{\rm mol} \rangle 
+ \int \! d^3R \left[  
\left( \nu - \frac{g^2}{m \lambda} \right) \langle \phi^\dagger \phi \rangle
+ \frac{1}{m\lambda} \langle \Phi^\dagger \Phi \rangle \right] ,
\label{energy-RM}
\end{eqnarray}
%-----------------
where $\langle V \rangle$ is the energy associated with the 
external potential,
%-----------------
\begin{equation}
\langle V \rangle = \int \! d^3R \, V(\bm{R}) 
\langle \psi_1^\dagger \psi_1 + \psi_2^\dagger \psi_2
                 + 2 \phi^\dagger \phi \rangle ,
\label{V-ave}
\end{equation}
%-----------------
and $\langle T_{\rm mol} \rangle$ is the kinetic energy 
of the molecules:
%-----------------
\begin{equation}
\langle T_{\rm mol} \rangle = 
\frac{1}{4m} \int \! d^3R \, \langle \nabla \phi^\dagger \cdot \nabla \phi \rangle .
\label{T-mol}
\end{equation}
%-----------------

%%%%%%%%%%%%%%%%%%%%%%%%%%%%%%%%%%%%%%%%%%%%%%%%%%%%%%%%%%%%%%%%%%%%%%%%%%
\section{Local pair density}
\label{sec:pairdensity}
%%%%%%%%%%%%%%%%%%%%%%%%%%%%%%%%%%%%%%%%%%%%%%%%%%%%%%%%%%%%%%%%%%%%%%%%%%

\begin{figure}[t]
\centerline{\includegraphics*[height=6cm,angle=0,clip=true]{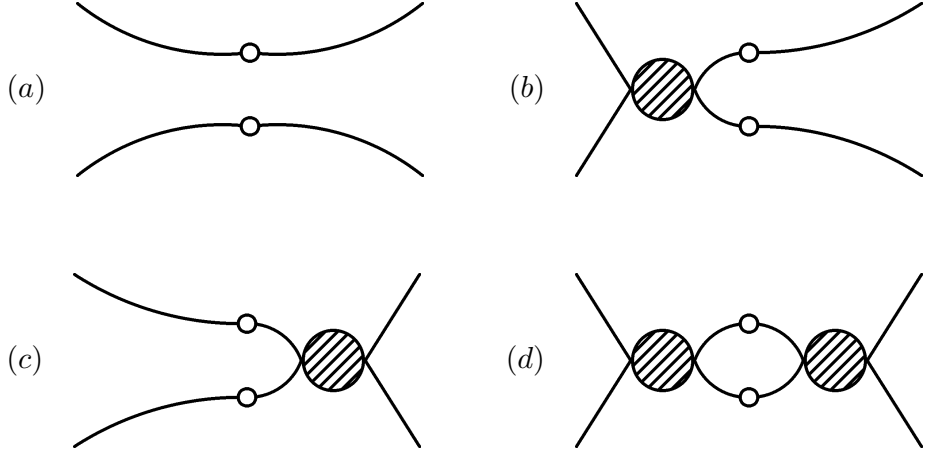}}
\vspace*{0.0cm}
\caption{The diagrams for the matrix element of the bilocal operator 
$\psi_1^\dagger \psi_1(\bm{R} -  \mbox{$\frac12$} \bm{r}) 
\psi_2^\dagger \psi_2(\bm{R} + \mbox{$\frac12$} \bm{r})$
between 2-atom scattering states.
The open dots represent the operators.
}
\label{fig:<nnbi>}
\end{figure}

The Tan relation in Eq.~(\ref{pairdensity-Tan}) expresses the contact 
density ${\cal C}$ in terms of the local pair density 
$N_{\rm pair}(\bm{R},s)$, which is the number of pairs of atoms 
in a ball of volume $\frac 43 \pi s^3$ centered at $\bm{R}$.
To obtain the analogous relation in the Resonance Model, 
we consider the OPE of the number density operators
$\psi_1^\dagger \psi_1(\bm{R} - \mbox{$\frac12$} \bm{r})$ and 
$\psi_2^\dagger \psi_2(\bm{R} + \mbox{$\frac12$} \bm{r})$.
The matrix elements of the product of these operators 
between 2-atom scattering states
can be represented diagramatically by the sum of the four diagrams 
in Fig.~\ref{fig:<nnbi>}.  The three diagrams in 
Figs.~\ref{fig:<nnbi>}(b,c,d) involve integrals over the 4-momenta 
of atoms.  The integrated momenta run through the lines that 
connect the operator vertices to the blobs.  Wilson coefficients 
that are not analytic at $\bm{r} = 0$ must come from the 
large-momentum regions of those integrals.  The corresponding 
operators can be deduced by shrinking the lines carrying the 
large momentum to a point.  
Shrinking those lines in the diagram in 
Fig.~\ref{fig:<nnbi>}(d) produces the 
operator vertices in Fig.~\ref{fig:linepoint} in the combination 
that corresponds to the operator $\Phi^\dagger \Phi(\bm{R})$.
The Wilson coefficient of this operator is calculated
in subsection~\ref{sec:me-n1n2} of the Appendix.  
It is proportional to $1/r^2$, 
so it is singular at $\bm{r} = 0$.  For the diagrams in 
Fig.~\ref{fig:<nnbi>}(b) and (c), shrinking the lines 
carrying large momentum produces operator vertices that 
correspond to the operators 
$\psi_1^\dagger \psi_2^\dagger \Phi(\bm{R})$ and
$\Phi^\dagger \psi_1 \psi_2(\bm{R})$, respectively.
The Wilson coefficient of these operators are 
proportional to $1/r$. 
Thus the most singular term in the OPE is 
%-----------------
\begin{eqnarray}
\psi_1^\dagger \psi_1(\bm{R} - \mbox{$\frac12$} \bm{r}) 
\psi_2^\dagger \psi_2(\bm{R} + \mbox{$\frac12$} \bm{r}) 
\longrightarrow  
\frac{1}{16 \pi^2 r^2} 
\Phi^\dagger \Phi(\bm{R}) .
\label{psi4-ope}
\end{eqnarray}
%-----------------

%%%%%%%%%%%%%%%%%%%%%%%%%%%%%%%%%%%%%%%%%%%%%%%%%%
%%%%%%%%%%%%%%%%%%%%%%%%%%%%%%%%%%%%%%%%%%%%%%%%%%

We can use the OPE in Eq.~(\ref{psi4-ope}) to obtain a simple 
interpretation of the contact density ${\cal C}$.
It can be expressed in the form
%-----------------
\begin{eqnarray}
\psi_1^\dagger \psi_1(\bm{R} + \bm{r}) 
\psi_2^\dagger \psi_2(\bm{R} + \bm{r}') 
\longrightarrow  \frac{1}{16 \pi^2 |\bm{r}'-\bm{r}|^2}
\Phi^\dagger \Phi(\bm{R}) ,
\label{psi4-ope:sing}
\end{eqnarray}
%-----------------
where on the right side we have kept only the most 
singular term as $\bm{r}' \to \bm{r}$.
If we integrate the left side of Eq.~(\ref{psi4-ope:sing})
over both $\bm{r}$ and $\bm{r}'$ inside the ball of 
volume $\frac43 \pi s^3$ centered at $\bm{R}$, we obtain an operator 
that in the absence of interactions would give the product 
of the number of atoms with spin 1 
and the number of atoms with spin 2 in that ball. 
Thus the expectation value $N_{\rm pair}(\bm{R}, s)$ of 
this operator can be interpreted as the number of pairs of atoms 
with spins 1 and 2 inside that ball.  
Integrating the right side of Eq.~(\ref{psi4-ope:sing})
over $\bm{r}$ and $\bm{r}'$ inside that same ball
and taking the expectation value, 
we obtain the limiting behavior of $N_{\rm pair}(\bm{R}, s)$
as $s \to 0$:
%-----------------
\begin{eqnarray}
N_{\rm pair}(\bm{R}, s) \longrightarrow
\frac{s^4}{4} \langle \Phi^\dagger \Phi(\bm{R}) \rangle.
\label{Npair-0}
\end{eqnarray}
%-----------------
Since $\langle \Phi^\dagger \Phi(\bm{R}) \rangle$ 
is the contact density ${\cal C}$,
the Tan relation in Eq.~(\ref{pairdensity-Tan})
is satisfied without modification in the Resonance Model.

%%%%%%%%%%%%%%%%%%%%%%%%%%%%%%%%%%%%%%%%%%%%%%%%%%%%%%%%%%%%%%%%%%%%%%%%%%
\section{Adiabatic relations}
\label{sec:adiabatic}
%%%%%%%%%%%%%%%%%%%%%%%%%%%%%%%%%%%%%%%%%%%%%%%%%%%%%%%%%%%%%%%%%%%%%%%%%%

Tan's adiabatic relation in Eq.~(\ref{adiabatic-Tan})
gives the rate of change in the energy 
$E$ of a state due to a change in the interaction parameter $a$. 
In the Resonance Model, there are three interaction parameters: 
$\nu$, $g$, and $\lambda$.
According to the expressions for these parameters in 
Eq.~(\ref{param-phys:RM}), $\nu$ can be changed 
experimentally by varying the magnetic field while 
$g$ and $\lambda$ are essentially constants.
Nevertheless, we will consider the effects of changing all 
three parameters.

The Feynman-Hellman theorem for variations in the parameter 
$\nu$ is
%-----------------
\begin{equation}
\partial E/\partial\nu = 
\int \! d^3R \, \langle \partial{\cal H}/\partial \nu \rangle  .
\label{E-nu:FH}
\end{equation}
%-----------------
There are analogous equations for the variations in the 
parameters $g$ and $\lambda$.
Since the Resonance Model is renormalizable, the Hamiltonian 
density operator ${\cal H} = {\cal T} + {\cal I} + {\cal V}$ 
given by Eqs.~(\ref{TIV:RM}) has ultraviolet-finite matrix elements.
More explicitly, if Eqs.~(\ref{param0:RM}) are used to
eliminate the bare parameters $\nu_0$, $g_0$, and $\lambda_0$ 
in favor of the renormalized parameters $\nu$, $g$, and 
$\lambda$, the matrix elements of ${\cal H}$ have finite limits as 
$\Lambda \to \infty$, where $\Lambda$ is the ultraviolet cutoff.
Since ${\cal H}$ has ultraviolet-finite matrix elements 
for any values of $\nu$, $g$, and $\lambda$, its derivatives
$(\partial/\partial \nu){\cal H}$, $
(\partial/\partial g){\cal H}$, and 
$(\partial/\partial \lambda){\cal H}$ must also have
ultraviolet-finite matrix elements.  The partial derivatives 
with the ultraviolet cutoff $\Lambda$ held fixed can be 
evaluated by using Eqs.~(\ref{param0:RM}) for the 
bare parameters.  The expressions for the partial derivatives
are simplest if they are expressed in terms of the field $\phi$ 
and the composite operator $\Phi$ defined in Eq.~(\ref{Phi-def}):
%-----------------
\begin{subequations}
\begin{eqnarray}
\nu \frac{\partial \ }{\partial \nu} {\cal H} &=& 
\nu \phi^\dagger \phi  ,
\label{H-nu}
\\
g \frac{\partial \ }{\partial g} {\cal H} &=& 
- \frac{2g^2}{m \lambda} \phi^\dagger \phi 
+ \frac{g}{m \lambda} 
\left( \phi^\dagger \Phi + \Phi^\dagger \phi \right) ,
\label{H-g}
\\
\lambda \frac{\partial \ }{\partial \lambda} {\cal H} &=& 
\frac{g^2}{m \lambda} \phi^\dagger \phi 
- \frac{g}{m \lambda} 
\left( \phi^\dagger \Phi + \Phi^\dagger \phi \right)
+ \frac{1}{m \lambda} \Phi^\dagger \Phi .
\label{H-lambda}
\end{eqnarray}
\label{H-param}
\end{subequations}
%-----------------
The right sides have been expressed as linear combinations 
of the operators $\phi^\dagger \phi$, 
$\phi^\dagger \Phi + \Phi^\dagger \phi$, and $\Phi^\dagger \Phi$,
with coefficients that are functions of renormalized parameters.
Since the three partial derivatives of ${\cal H}$ 
in Eqs.~(\ref{H-param}) must have
ultraviolet-finite matrix elements, these three operators
must also have ultraviolet-finite matrix elements. 

The adiabatic relations for the  Resonance Model can be
obtained by taking the expectation values of the 
partial derivatives in Eqs.~(\ref{H-param}) 
and then integrating over all space:
%-----------------
\begin{subequations}
\begin{eqnarray}
\nu \frac{\partial \ }{\partial \nu} E &=& 
\nu \int \! d^3R \, \langle \phi^\dagger \phi \rangle  ,
\label{adiabatic-RM:nu}
\\
g \frac{\partial \ }{\partial g} E &=& 
\int \! d^3R \left(
- \frac{2g^2}{m\lambda} \langle \phi^\dagger \phi \rangle 
+ \frac{g}{m\lambda} \langle \phi^\dagger \Phi 
+ \Phi^\dagger \phi \rangle \right) ,
\label{adiabatic-RM:g}
\\
\lambda \frac{\partial \ }{\partial \lambda} E &=& 
\int \! d^3R 
\left(  \frac{g^2}{m\lambda} \langle \phi^\dagger \phi \rangle 
-  \frac{g}{m\lambda} \langle \phi^\dagger \Phi + \Phi^\dagger \phi \rangle 
+ \frac{1}{m\lambda}\langle \Phi^\dagger \Phi \rangle \right) .
\label{adiabatic-RM:lambda}
\end{eqnarray}
\label{adiabatic-RM}
\end{subequations}
%-----------------

%%%%%%%%%%%%%%%%%%%%%%%%%%%%%%%%%%%%%%%%%%%%%%%%%%%%%%%%%%%%%%%%%%%%%%%%%%
\section{Pressure relation}
\label{sec:pressure}
%%%%%%%%%%%%%%%%%%%%%%%%%%%%%%%%%%%%%%%%%%%%%%%%%%%%%%%%%%%%%%%%%%%%%%%%%%

For a homogeneous system, the external potential is a constant:
$V(\bm{r}) = - \mu$, where $\mu$ is the chemical potential.
The thermodynamic properties of the system are determined by the free 
energy density (or thermodynamic potential density)
${\cal F}$.  It can be expressed in terms 
of the trace of the partition function:
%-----------------
\begin{equation}
\exp (- \beta {\cal F} V) = 
{\rm Tr} \exp (- \beta \mbox{$\int$} d^3R \, {\cal H}) ,
\label{F-def}
\end{equation}
%-----------------
where $\beta= 1/T$ and $V$ is the volume. 
The free energy density ${\cal F}$ is a function of 
$T$, $\mu$, and the interaction parameters $\nu$, $g$, and 
$\lambda$.  The parameter $\nu$ has dimensions of energy.
The mass $m$, together with $\hbar = 1$, can be used as a 
conversion constant to obtain energy scales 
$1/(m \lambda^2)$ and $g^4/m$ associated with the two other
interaction parameters.  The free energy density ${\cal F}$ 
has dimensions of energy per volume, which can be expressed as 
(energy)$^{5/2}$ up to factors of $m$ and $\hbar$.  
Thus dimensional analysis requires that the free energy density
satisfy
%-----------------
\begin{equation}
\left[ T \frac{\partial \ }{\partial T} 
+ \mu \frac{\partial \ }{\partial \mu} 
+ \nu \frac{\partial \ }{\partial \nu} 
+ \frac14 g \frac{\partial \ }{\partial g}
- \frac12 \lambda \frac{\partial \ }{\partial \lambda} 
\right] {\cal F} = \frac52 {\cal F} .
\label{F-diman:RM}
\end{equation}
%-----------------

The logarithmic derivatives on the left side of Eq.~(\ref{F-diman:RM})
can be determined by applying them to the logarithms of both sides 
of Eq.~(\ref{F-def}).  
In the case of the interaction parameters $\nu$, $g$, and $\lambda$,
the logarithmic derivatives of ${\cal F}$ are simply the 
expectation values of the logarithmic derivatives of the 
Hamiltonian density given in Eqs.~(\ref{H-param}).
In the case of the temperature and the chemical potential, the 
logarithmic derivatives of ${\cal F}$ are
%-----------------
\begin{subequations}
\begin{eqnarray}
T \frac{\partial \ }{\partial T} {\cal F} &=& 
{\cal F} - {\cal E} + \mu n  ,
\\
\mu \frac{\partial \ }{\partial \mu} {\cal F} &=& - \mu n   ,
\end{eqnarray}
\end{subequations}
%-----------------
where $n$ is the expectation value of the total number density:
%-----------------
\begin{equation}
n = \langle \psi_1^\dagger \psi_1 + \psi_2^\dagger \psi_2
+ 2 \phi^\dagger \phi \rangle .
\label{n-def}
\end{equation}
%-----------------
Because the Hamiltonian density ${\cal H}$ includes the term
${\cal V}$ in Eq.~(\ref{V:RM}), it is actually the 
grand Hamiltonian density for the grand canonical ensemble. 
Its expectation value is
$\langle {\cal H} \rangle = {\cal E} - \mu n$, where 
${\cal E}$ is the energy density. 
Because the system is homogeneous, the pressure is simply 
${\cal P} = - {\cal F}$.
Combining together all the terms in Eq.~(\ref{F-diman:RM}), we obtain
%-----------------
\begin{equation}
{\cal E} = \frac32 {\cal P}
+ \left( \nu - \frac{g^2}{m \lambda} \right) \langle \phi^\dagger \phi \rangle
+ \frac{3g}{4 m \lambda} \langle \phi^\dagger \Phi + \Phi^\dagger \phi \rangle
- \frac{1}{2 m \lambda} \langle \Phi^\dagger \Phi \rangle .
\label{pressure-RM}
\end{equation}
%-----------------
This is the {\it pressure relation} for the Resonance Model.

%%%%%%%%%%%%%%%%%%%%%%%%%%%%%%%%%%%%%%%%%%%%%%%%%%%%%%%%%%%%%%%%%%%%%%%%%%
\section{Virial theorem}
\label{sec:virial}
%%%%%%%%%%%%%%%%%%%%%%%%%%%%%%%%%%%%%%%%%%%%%%%%%%%%%%%%%%%%%%%%%%%%%%%%%%

If the trapping potential $V(\bm{R})$ has simple behavior 
under scaling the coordinates $\bm{R} = (X,Y,Z)$,
the virial theorem gives simple
relations between various contributions to the energy.
The most important case for cold atoms is a harmonic 
trapping potential.  For simplicity, we consider only 
the case of an isotropic harmonic potential:
%-----------------
\begin{equation}
V(\bm{R}) = \frac12 m \omega^2 \bm{R}^2 .
\label{V-R}
\end{equation}
%-----------------
The generalization to the case of an anisotropic harmonic 
potential is straightforward.

The virial theorem can be derived simply from dimensional 
analysis.  Let us assume that the state of the system 
can be specified by a set of integer quantum numbers.
The energy $E = \int d^3r \langle {\cal H} \rangle$ for any such state 
must be a function of the frequency $\omega$ and the 
interaction parameters $\nu$, $g$, and $\lambda$.  
Since $E$ has dimensions of energy, 
dimensional analysis requires that it satisfy
%-----------------
\begin{equation}
\left[ \omega \frac{\partial \ }{\partial \omega} 
+ \nu \frac{\partial \ }{\partial \nu} 
+ \frac14 g \frac{\partial \ }{\partial g}
- \frac12 \lambda \frac{\partial \ }{\partial \lambda} 
\right] E = E .
\label{E-diman:RM}
\end{equation}
%-----------------
If we consider the system at a nonzero temperature $T$,
the differential operator in Eq.~(\ref{E-diman:RM}) 
would also include a term $T(\partial/\partial T)$.

The action of the differential operator in Eq.~(\ref{E-diman:RM}) 
on the Hamiltonian density can be obtained by inserting 
the partial derivatives in Eqs.~(\ref{H-param}):
%-----------------
\begin{eqnarray}
\left[ \omega \frac{\partial \ }{\partial \omega} 
+ \nu \frac{\partial \ }{\partial \nu} 
+ \frac14 g \frac{\partial \ }{\partial g}
- \frac12 \lambda \frac{\partial \ }{\partial \lambda} 
\right] {\cal H} = 
2 V(\bm{R}) \left( \psi_1^\dagger \psi_1 + \psi_2^\dagger \psi_2
                 + 2 \phi^\dagger \phi \right)
\nonumber
\\
+ \left( \nu - \frac{g^2}{m \lambda} \right) \phi^\dagger \phi
+ \frac{3g}{4 m \lambda} 
\left( \phi^\dagger \Phi + \Phi^\dagger \phi \right)
- \frac{1}{2 m \lambda} \Phi^\dagger \Phi .
\label{H-diman}
\end{eqnarray}
%-----------------
The virial theorem for the Resonance Model is obtained 
by taking the expectation value,
integrating over space, and setting this equal to $E$:
%-----------------
\begin{eqnarray}
E &=& 2 \langle V \rangle
+ \int \! d^3R \left[ 
\left( \nu - \frac{g^2}{m \lambda} \right) \langle \phi^\dagger \phi \rangle
+ \frac{3g}{4 m \lambda} \langle \phi^\dagger \Phi + \Phi^\dagger \phi \rangle
- \frac{1}{2 m \lambda} \langle \Phi^\dagger \Phi \rangle \right] ,
\label{virial-RM}
\end{eqnarray}
%-----------------
where $\langle V \rangle$ is the energy from the 
external potential given by Eq.~(\ref{V-ave}).

%%%%%%%%%%%%%%%%%%%%%%%%%%%%%%%%%%%%%%%%%%%%%%%%%%%%%%%%%%%%%%%%%%%%%%%%%%
\section{Inelastic two-body loss rate}
\label{sec:lossrate}
%%%%%%%%%%%%%%%%%%%%%%%%%%%%%%%%%%%%%%%%%%%%%%%%%%%%%%%%%%%%%%%%%%%%%%%%%%

If the atoms have other spin states with lower energy
than the spin states $\sigma= 1,2$ of interest, a pair of atoms 
with spins 1 and 2 may be able to scatter inelastically into the 
lower spin states.  The molecular state responsible for the 
Feshbach resonance may also be able to decay into pairs of atoms
in lower spin states.  If the lower spin states interact
weakly with the spin states 1 and 2, the primary effects 
of transitions to the lower spin states is the disappearance 
of atoms from the spin states 1 and 2 and the disappearance 
of the Feshbach molecule.  The inclusive 
effects of the lower spin states on states near the threshold 
for the spin states 1 and 2 can be taken into account through
anti-hermitian terms in the Hamiltonian density.
These terms must be anti-hermitian to account for the loss of
probability from transitions to the lower spin states, 
which are not taken into account as explicit degrees of freedom.

If the energy gap between the
lower spin states and the spin states 1 and 2 
is much larger than the energies of interest, 
the anti-hermitian terms in the Hamiltonian density
can be chosen to be local operators.
The rate of disappearance of molecules 
through their decays into pairs of atoms in lower spin states
is proportional to the number density of the molecules.
The resulting effects can be taken into account through the operator 
$\phi^\dagger \phi$ with a negative imaginary 
coefficient.  The rate of disappearance of pairs of atoms 
through inelastic scattering is proportional to the local  
density of pairs of atoms.  The resulting effects can be taken 
into account through the operator
$\psi_1^\dagger \psi_2^\dagger \psi_1 \psi_2$ 
with a negative imaginary coefficient.  If the decay of the
molecule and the inelastic scattering of a pair of atoms 
can produce the same final states, the resulting interference 
effects can be taken into account through the operators
$\phi^\dagger \psi_1 \psi_2$ and
$\psi_1^\dagger \psi_2^\dagger \phi$.
The large energy gap $E_{\rm gap}$ is essential for allowing
the effects of transitions into the lower spin states
to be taken into account through local operators.
The atoms created by the decay of a molecule or by the 
inelastic scattering of atoms with spins 1 and 2
will emerge with large momentum given
approximately by $(m E_{\rm gap})^{1/2}$.
By the uncertainty principle, a localized wave packet 
corresponding to such an atom can be traced back to a region 
of origin whose size is approximately 
$\hbar/(m E_{\rm gap})^{1/2}$.  But the regions of origin of the 
two atoms must also be separated by a distance
at most of order $\hbar/(m E_{\rm gap})^{1/2}$.
The reason for this is that the interactions that change 
the spin state of the atom from 1 or 2 to a lower spin state 
must also deliver a momentum kick of order $(m E_{\rm gap})^{1/2}$.
This large momentum can be transferred between the two 
final state atoms only if they are close together.
Even if it is transferred though an intermediate
atom with spin 1 or 2, that atom must be far off its energy shell 
and therefore can propagate only over a short distance.
As long as $\hbar/(m E_{\rm gap})^{1/2}$
is small compared to the length scales that are 
described explicitly by the field theory, the effects of 
the transitions to lower spin states can be taken into account 
systematically through local operators.

The operators $\psi_1^\dagger \psi_2^\dagger \psi_1 \psi_2$, 
$\phi^\dagger \psi_1 \psi_2+\psi_1^\dagger \psi_2^\dagger \phi$, 
and $\phi^\dagger \phi$
required to take into account the leading effects 
of the lower spin states are ones that appear already 
in the interaction term in the Hamiltonian density
given in Eq.~(\ref{I:RM}).  
To give anti-hermitian terms in the Hamiltonian density, 
these operators must be multiplied by imaginary coefficients.
Thus the leading effects of the lower spin states can be
taken into account by adding negative imaginary parts to the bare
parameters $\nu_0$, $g_0$, and $\lambda_0$.
We now consider the effect of the nonhermitian terms in the 
Hamiltonian.
The expectation value of those terms can be expressed as
$-i \Gamma/2$, where
%-----------------
\begin{eqnarray}
\Gamma &\approx& 
- 2 \, \int \! d^3R \, \left[
{\rm Im} \nu_0 \langle \phi^\dagger \phi \rangle
+ \frac{{\rm Im} g_0}{m}
	\langle \phi^\dagger \psi_1 \psi_2^{(\Lambda)} 
	+ \psi_1^\dagger \psi_2^\dagger \phi^{(\Lambda)} \rangle
+ \frac{{\rm Im} \lambda_0}{m} 
\langle \psi_1^\dagger \psi_2^\dagger \psi_1 \psi_2^{(\Lambda)} \rangle \right] .
\label{Gamma-RM:bare}
\end{eqnarray}
%-----------------
The leading effect of these terms on a state with definite energy $E$ 
is to change its time-dependence from
$\exp(-i E t/\hbar)$ to $\exp(-i (E - i \Gamma/2)t/\hbar)$.
The probability in such a state decreases with time at the rate 
$\Gamma/\hbar$.  Thus $\Gamma$ is the rate at which 
atoms with spins 1 and 2 and Feshbach molecules are depleted by 
transitions to pairs of atoms in lower spin states.

The loss rate in Eq.~(\ref{Gamma-RM:bare}) can be expressed 
in terms of expectation values 
of operators with ultraviolet-finite matrix elements
by eliminating $\psi_1 \psi_2$ in favor of the operator $\Phi$ 
defined in Eq.~(\ref{Phi-def}).  Since the coefficients  
of these expectation values must also 
be ultraviolet finite, they are expressible in terms of the 
renormalized parameters $\nu$, $g$, and $\lambda$.
Using Eqs.~(\ref{param0:RM}) to eliminate the bare parameters 
and expanding to first order in the imaginary parts of the 
renormalized parameters, we obtain
%-----------------
\begin{eqnarray}
\Gamma &\approx& 
- 2 \, \int \! d^3R \left[ 
\left( {\rm Im} \nu - \frac{2g}{m \lambda} {\rm Im} g  
	+ \frac{g^2}{m \lambda^2} {\rm Im} \lambda \right) 
	\langle \phi^\dagger \phi \rangle
\right.
\nonumber
\\
&& \hspace{2cm} \left.
+ \left( \frac{1}{m \lambda} {\rm Im} g  
	- \frac{g}{m \lambda^2} {\rm Im} \lambda \right) 
	\langle \phi^\dagger \Phi + \Phi^\dagger \phi \rangle
+ \frac{1}{m\lambda^2} {\rm Im} \lambda
	\langle \Phi^\dagger \Phi \rangle \right] .
\label{Gamma-RM}
\end{eqnarray}
%-----------------
This is the generalization in the Resonance Model
of the Tan relation for the {\it inelastic 2-body loss rate}.

%%%%%%%%%%%%%%%%%%%%%%%%%%%%%%%%%%%%%%%%%%%%%%%%%%%%%%%%%%%%%%%%%%%%%%%%%%
\section{Constraints for Large Scattering Length}
\label{sec:constraints}
%%%%%%%%%%%%%%%%%%%%%%%%%%%%%%%%%%%%%%%%%%%%%%%%%%%%%%%%%%%%%%%%%%%%%%%%%%

The Tan relations in Eqs.~(\ref{tail-Tan})--(\ref{Gam-Tan}) 
express various properties of an arbitrary state 
consisting of atoms with two spin states and a large 
scattering length in terms of the contact $C$ or the 
contact density ${\cal C}$.  
These relations hold exactly in the Zero-Range Model.
The contact density operator in this model was identified 
in Ref.~\cite{Braaten:2008uh} to be
$\lambda_0^2 \psi_1^\dagger \psi_2^\dagger \psi_1 \psi_2$.
In the Resonance Model, the generalizations of the Tan relations
derived in Sections~\ref{sec:tail}--\ref{sec:lossrate}
involve the expectation values of three distinct local operators:
$\Phi^\dagger \Phi$, $\phi^\dagger \Phi + \Phi^\dagger \phi$, 
and $\phi^\dagger \phi$, where $\Phi$ is the 
composite operator defined in Eq.~(\ref{Phi-def}).
The contact density operator is identified in 
Eq.~(\ref{contact-RM}) to be $\Phi^\dagger \Phi$.
The same operator appears in the energy relation in
Eq.~(\ref{energy-RM}) and in the local pair density 
in Eq.~(\ref{Npair-0}).
The three adiabatic relations in Eqs.~(\ref{adiabatic-RM}),
the virial theorem in Eq.~(\ref{virial-RM}),
the pressure relation in Eq.~(\ref{pressure-RM}),
and the 2-body inelastic loss rate in Eq.~(\ref{Gamma-RM})
involve various linear combinations of the expectation values 
of the three local operators.

The scattering length is the only interaction parameter 
in the Zero-Range Model.
The three parameters of the Resonance Model provide three distinct 
length scales: $|m \nu|^{-1/2}$, $g^{-2}$ and $|\lambda|$.
There are two limits in which the scattering length 
given by Eq.~(\ref{a:RM}) becomes arbitrarily large compared 
to the other two length scales.  In particular, it becomes large 
compared to the effective range given by Eq.~(\ref{r:RM}).
One of these limits is $\lambda \to \pm \infty$ with 
$\nu$ and $g$ fixed, in which case $a \to \lambda/(4 \pi)$.
The other limit is $\nu \to 0$ with $g$ and $\lambda$ fixed,
in which case $a \to - g^2/(4 \pi m \nu)$.
In either of these two limits, the universality of systems 
with large scattering length requires the Resonance Model 
to reduce to the Zero-Range Model.  The generalized Tan relations 
must therefore reduce to the original Tan relations.
We will find that this imposes constraints on matrix elements 
of $\phi^\dagger \phi$ and $\phi^\dagger \Phi + \Phi^\dagger \phi$.

\subsection{$\bm{\lambda \to \pm \infty}$ with $\bm{\nu}$ and $\bm{g}$ fixed}

We first consider the case of a large scattering length $a$ obtained 
by increasing $|\lambda|$ with $\nu$ and $g$ fixed.
The adiabatic relation in Eq.~(\ref{adiabatic-RM:lambda}) can be expressed 
as an adiabatic relation for variations in $a$: 
%-----------------
\begin{equation}
\frac{d \ }{d a} E = 
\frac{4 \pi}{m \lambda^2} \int \! d^3R 
\left( g^2 \langle \phi^\dagger \phi \rangle 
-  g \langle \phi^\dagger \Phi + \Phi^\dagger \phi \rangle 
+ \langle \Phi^\dagger \Phi \rangle \right) .
\label{adiabatic-RM:a}
\end{equation}
%-----------------
When $\lambda$ is sufficiently large, this must reduce to 
Tan's adiabatic relation in Eq.~(\ref{adiabatic-Tan}), 
which can be expressed as
%-----------------
\begin{equation}
\frac{d \ }{d a} E = 
\frac{4 \pi}{m \lambda^2} \int \! d^3R \, \langle \Phi^\dagger \Phi \rangle 
+ O(1/\lambda^3) .
\label{dEda-Phi:a}
\end{equation}
%-----------------
This implies a constraint on the matrix elements in 
Eq.~(\ref{adiabatic-RM:a}) as $\lambda \to \pm \infty$:
%-----------------
\begin{equation}
g^2 \langle \phi^\dagger \phi \rangle 
-  g \langle \phi^\dagger \Phi + \Phi^\dagger \phi \rangle 
=  O(1/\lambda) .
\label{constraint-adiabatic:a}
\end{equation}
%-----------------
We can obtain an independent constraint on the matrix elements 
by using the virial theorem, which for the Resonance Model
is given in Eq.~(\ref{virial-RM}).
When $\lambda$ is sufficiently large, this must reduce to 
Tan's virial theorem in Eq.~(\ref{virial-Tan}), 
which can be expressed as
%-----------------
\begin{equation}
E = 2 \langle V \rangle
- \frac{1}{2 m \lambda} \int \! d^3R \, \langle \Phi^\dagger \Phi \rangle 
+ O(1/\lambda^2) .
\label{virial-RM:a}
\end{equation}
%-----------------
This implies a constraint on the matrix elements in 
Eq.~(\ref{virial-RM}) as $\lambda \to \pm \infty$:
%-----------------
\begin{equation}
\left( \nu - \frac{g^2}{m \lambda} \right) \langle \phi^\dagger \phi \rangle
+ \frac{3g}{4 m \lambda} \langle \phi^\dagger \Phi + \Phi^\dagger \phi \rangle
= O(1/\lambda^2) .
\label{constraint-virial:a}
\end{equation}
%-----------------

If $\nu \neq 0$, the constraints in Eqs.~(\ref{constraint-adiabatic:a})
and (\ref{constraint-virial:a}) imply
%-----------------
\begin{subequations}
\begin{eqnarray}
\langle \phi^\dagger \phi \rangle &=& O(1/\lambda^2),
\\
\langle \phi^\dagger \Phi + \Phi^\dagger \phi \rangle &=& O(1/\lambda) .
\end{eqnarray}
\label{constraints:a}
\end{subequations}
%-----------------
These constraints are consistent with the decoupling of the 
molecular field $\phi$ in the limit $\lambda \to \pm \infty$.
Matrix elements of operators involving the field $\phi$ 
are suppressed by a factor of $1/\lambda$ 
for every factor of $\phi$ in the operator.

\subsection{$\bm{\nu \to 0}$ with $\bm{g}$ and $\bm{\lambda}$ fixed}

We next consider the case of a large scattering length $a$ obtained 
by decreasing $|\nu|$ with $g$ and $\lambda$ fixed.
According to Eq.~(\ref{nu-phys:RM}),
this can be achieved by tuning the magnetic field to the position
$B_0$ of the Feshbach resonance.
The adiabatic relation in Eqs.~(\ref{adiabatic-RM:nu}) can be expressed 
as an adiabatic relation for variations in $a$: 
%-----------------
\begin{equation}
\frac{d \ }{d a} E = 
\frac{4 \pi m \nu^2}{g^2} \int \! d^3R \, 
\langle \phi^\dagger \phi \rangle .
\label{adiabatic-RM:b}
\end{equation}
%-----------------
When $\nu$ is sufficiently small, this must reduce to 
Tan's adiabatic relation in Eq.~(\ref{adiabatic-Tan}), 
which can be expressed as
%-----------------
\begin{equation}
\frac{d \ }{d a} E = 
\frac{4 \pi m \nu^2}{g^4} \int \! d^3R 
\, \langle \Phi^\dagger \Phi \rangle 
+ O(\nu^3) .
\label{dEda-Phi:b}
\end{equation}
%-----------------
This implies that in the limit $\nu \to 0$, the expectation value 
of $\phi^\dagger \phi$ can be expressed in terms of the contact 
density:
%-----------------
\begin{equation}
g^2 \langle \phi^\dagger \phi \rangle 
= \langle \Phi^\dagger \Phi \rangle + O(\nu) .
\label{constraint-adiabatic:b}
\end{equation}
%-----------------

We can obtain an independent constraint on the matrix elements 
by using the virial theorem, which for the Resonance Model
is given in Eq.~(\ref{virial-RM}).
When $\nu$ is sufficiently small, this must reduce to 
Tan's virial theorem in Eq.~(\ref{virial-Tan}), 
which can be expressed as
%-----------------
\begin{equation}
E = 2 \langle V \rangle
+ \frac{\nu}{2 g^2}  \int \! d^3R \, \langle \Phi^\dagger \Phi \rangle 
+ O(\nu^2) .
\label{constraint:a}
\end{equation}
%-----------------
This implies a constraint on the matrix elements in 
Eq.~(\ref{virial-RM}) as $\nu \to 0$:
%-----------------
\begin{equation}
\left( \nu - \frac{g^2}{m \lambda} \right) \langle \phi^\dagger \phi \rangle
+ \frac{3g}{4 m \lambda} \langle \phi^\dagger \Phi + \Phi^\dagger \phi \rangle
- \frac{1}{2 m \lambda} \langle\Phi^\dagger \Phi \rangle
= \frac{\nu}{2 g^2} \langle \Phi^\dagger \Phi \rangle
+ O(\nu^2) .
\label{constraint-virial:b}
\end{equation}
%-----------------
Combined with the constraint in Eq.~(\ref{constraint-adiabatic:b}),
this implies that the expectation value of 
$\phi^\dagger \Phi + \Phi^\dagger \phi$ can also
be expressed in terms of the contact density:
%-----------------
\begin{equation}
g \langle \phi^\dagger \Phi + \Phi^\dagger \phi \rangle = 
2 \langle \Phi^\dagger \Phi \rangle + O(\nu) .
\label{constraint:b}
\end{equation}
%-----------------
It also implies a constraint on the terms of order $\nu$
on the right sides of Eqs.~(\ref{constraint-adiabatic:b}) and 
(\ref{constraint:b}):
%-----------------
\begin{equation}
\left( g^2 \langle \phi^\dagger \phi \rangle 
	- \langle \Phi^\dagger \Phi \rangle \right) 
- \frac34 \left( 
g \langle \phi^\dagger \Phi + \Phi^\dagger \phi \rangle
	- 2 \langle \Phi^\dagger \Phi \rangle \right)
= \frac{\lambda m \nu}{2g^2} \langle \Phi^\dagger \Phi \rangle
+ O(\nu^2) .
\label{constraint:c}
\end{equation}
%-----------------

%%%%%%%%%%%%%%%%%%%%%%%%%%%%%%%%%%%%%%%%%%%%%%%%%%%%%%%%%%%%%%%%%%%%%%%%%%
\section{Effective Range Model}
\label{sec:ERM}
%%%%%%%%%%%%%%%%%%%%%%%%%%%%%%%%%%%%%%%%%%%%%%%%%%%%%%%%%%%%%%%%%%%%%%%%%%

If we set $\lambda = 0$ in the S-wave phase shift for the 
Resonance Model, which is given by Eq.~(\ref{delta0:RM}),
it reduces to
\begin{equation}
k \cot \delta_0(k) = 
\frac{4 \pi m \nu}{g^2} - \frac{4 \pi}{g^2}k^2 .
\label{delta0:ERM}
\end{equation}
This is the phase shift of the Effective Range Model in  
Eq.~(\ref{delta0:ars}),
with the scattering length and the effective range given by
\begin{subequations}
\begin{eqnarray}
a &=&  - \frac{g^2}{4 \pi m \nu}  ,
\label{a:ERM}
\\
r_s &=&  - \frac{8 \pi}{g^2}.
\label{r:ERM}
\end{eqnarray}
\label{ar:ERM}
\end{subequations}
Note that the effective range is negative.

In the simplest quantum field theory formulation 
of the Effective Range Model \cite{BKZ0709},
there are three quantum fields: fermionic fields $\psi_1$ and $\psi_2$
that annihilate atoms
and a bosonic field $\phi$ that annihilates a diatomic molecule.
The Hamiltonian density ${\cal H} = {\cal T} +{\cal I}+{\cal V}$
for the Effective Range Model is the sum of
the kinetic term in Eq.~(\ref{T:RM}), 
the external potential term  in Eq.~(\ref{V:RM}),  
and the interaction term 
%-----------------
\begin{equation}
{\cal I} =   \nu_0 \phi^\dagger \phi
+ \frac{g}{m} \left(\phi^\dagger \psi_1 \psi_2^{(\Lambda)} 
+ \psi_1^\dagger \psi_2^\dagger \phi^{(\Lambda)} \right) .
\label{I:ERM}
\end{equation}
%-----------------
To avoid ultraviolet divergences, an ultraviolet cutoff $\Lambda$
must be imposed on the momenta of virtual particles.
The only renormalization required in this model is an additive 
renormalization of the parameter $\nu_0$:
\begin{equation}
\nu_0 (\Lambda) = \nu + \frac{g^2 \Lambda}{2 \pi^2 m} .
\label{nu0:ERM}
\end{equation}
The Effective Range Model can also be formulated as a 
local quantum field theory with fermion fields $\psi_1$ and 
$\psi_2$ only and with a momentum-dependent interaction
\cite{BKZ0709}.  In this formulation of the model,
a rather intricate renormalization of the parameters 
is required to obtain the phase shift in Eq.~(\ref{delta0:ERM})
in the limit $\Lambda \to \infty$.

The generalized Tan relations for the Effective Range Model
can be derived by using the same methods that we used for the 
Resonance Model.  However they can be derived more easily 
by starting from the generalized Tan relations 
for the Resonance Model and taking the limit $\lambda \to 0$.  
This limit is not trivial, because some of the generalized
Tan relations for the Resonance Model involve factors of $1/\lambda$.
The limit $\lambda \to 0$ can be taken by expanding the composite 
operator $\Phi$ defined in Eq.~(\ref{Phi-def}) through first 
order in $\lambda$.  Using the expressions for the bare coupling 
constants in Eqs.~(\ref{lambda0:RM}) and (\ref{g0:RM})
to expand $\lambda_0$ and $g_0$ in powers of $\lambda$, we obtain
%-----------------
\begin{equation}
\Phi(\bm{R}) = g \phi(\bm{R}) 
+ \lambda \left[ \psi_1 \psi_2^{(\Lambda)}(\bm{R}) 
	+ \frac{g \Lambda}{2 \pi^2} \phi(\bm{R}) \right] + O(\lambda^2).
\label{Phi-lambda}
\end{equation}
%-----------------
We denote the composite operator defined by the coefficient of 
$\lambda$ by $\Phi'(\bm{R})$:
%-----------------
\begin{equation}
\Phi'(\bm{R}) = \psi_1 \psi_2^{(\Lambda)}(\bm{R})
+ \frac{g \Lambda}{2 \pi^2} \phi(\bm{R}) .
\label{Phi'-def}
\end{equation}
%-----------------

The contact $C$ in the Resonance Model is given in 
Eq.~(\ref{contact-RM}).  From the expansion in 
Eq.~(\ref{Phi-lambda}), we can see that the limit $\lambda \to 0$ 
can be taken simply by replacing $\Phi$ by $g \phi$.
Thus the contact in the Effective Range Model is
%-----------------
\begin{equation}
C = \int \! d^3R \,
g^2 \langle X | \phi^\dagger \phi(\bm{R}) | X \rangle  .
\label{contact-ERM}
\end{equation}
%-----------------

We proceed to enumerate the generalized Tan relations for the 
Effective Range Model:
\begin{enumerate}

\item
{\bf Tail of the momentum distribution}.
 This is given by Eq.~(\ref{tail-Tan}), where $C$ is the contact 
in Eq.~(\ref{contact-ERM}).

\item
{\bf Energy relation}.
The decomposition of the energy into
terms that are ultraviolet-finite
can be obtained by taking the limit $\lambda \to 0$ in 
Eq.~(\ref{energy-RM}):
%-----------------
\begin{eqnarray}
E &=& \langle V \rangle
+ \langle T_1^{\rm (sub)} \rangle
+ \langle T_2^{\rm (sub)} \rangle
+ \langle T_{\rm mol} \rangle
+ \int \! d^3R
\left( \nu \langle \phi^\dagger \phi \rangle
+ \frac{g}{m} \langle \phi^\dagger \Phi' +\Phi^{'\dagger} \phi \rangle \right),
\label{energy-ERM}
\end{eqnarray}
%-----------------
where $\Phi'$ is the composite operator defined by Eq.~(\ref{Phi'-def}).
The subtracted kinetic energy for a single spin state 
is obtained from Eq.~(\ref{T-sub:k}) by replacing $\Phi$ by $g \phi$:
%-----------------
\begin{equation}
\langle T_\sigma^{\rm (sub)} \rangle =
\frac{1}{2m} \int \! d^3R \,
\left( \langle \nabla \psi_\sigma^\dagger
       \cdot \nabla \psi_\sigma^{(\Lambda)} \rangle
     - \frac{g^2 \Lambda}{2 \pi^2} \langle \phi^\dagger \phi \rangle \right) .
\label{T-sub:ERM}
\end{equation}
%-----------------

\item
{\bf Local pair density}.
The contact density ${\cal C} = g^2 \langle \phi^\dagger \phi \rangle$
is related to the local pair density $N_{\rm pair}(\bm{R},s)$
by Eq.~(\ref{pairdensity-Tan}).

\item
{\bf Adiabatic relation}.
The adiabatic relation for variations in $\nu$ is the same as
in the Resonance Model, where it is given by
Eq.~(\ref{adiabatic-RM:nu}). The adiabatic relation for variations
in $g$ can be obtained from Eq.~(\ref{adiabatic-RM:g}) 
by taking the limit $\lambda \to 0$.
The adiabatic relations are therefore
%-----------------
\begin{subequations}
\begin{eqnarray}
\nu \frac{\partial \ }{\partial \nu} E &=&
\nu \int \! d^3R \, \langle \phi^\dagger \phi \rangle  ,
\label{adiabatic-ERM:nu}
\\
g \frac{\partial \ }{\partial g} E &=&
\frac{g}{m} \int \! d^3R 
\langle \phi^\dagger \Phi' + \Phi^{'\dagger} \phi \rangle .
\label{adiabatic-ERM:g}
\end{eqnarray}
\label{adiabatic-ERM}
\end{subequations}
%-----------------

\item
{\bf Pressure relation}.
The relation between the energy density ${\cal E}$
and the pressure ${\cal P}$ in a homogeneous system 
can be obtained from Eq.~(\ref{pressure-RM}) 
by taking the limit $\lambda \to 0$:
%-----------------
\begin{equation}
{\cal E} = \frac32 {\cal P}
+ \nu \langle \phi^\dagger \phi \rangle
+ \frac{g}{4 m} \langle \phi^\dagger \Phi' + \Phi^{'\dagger} \phi \rangle .
\label{pressure-ERM}
\end{equation}
%-----------------

\item
{\bf Virial theorem}.
The {\it virial theorem} for a system
with a harmonic trapping potential
can be obtained from Eq.~(\ref{virial-RM}) 
by taking the limit $\lambda \to 0$:
%-----------------
\begin{equation}
E = 2 \langle V \rangle
+ \int \! d^3R \left(
\nu \langle \phi^\dagger \phi \rangle
+ \frac{g}{4 m} \langle \phi^\dagger \Phi' + \Phi^{'\dagger} \phi \rangle
\right) .
\label{virial-ERM}
\end{equation}
%-----------------
The virial theorem for the Effective Range Model
has been derived previously by Werner \cite{Werner:0803}.
His result can be expressed in the form
%-----------------
\begin{equation}
E = 2 \langle V \rangle
- \frac{a}{2} \frac{\partial \ }{\partial a} E
- \frac{r_s}{2} \frac{\partial \ }{\partial r_s} E .
\label{virial-Werner}
\end{equation}
%-----------------
This is just the generalization of the dimensional analysis 
condition for the energy in Eq.~(\ref{E-diman})
for a model with two interaction parameters $a$ and $r_s$ 
with dimensions of length.
If we use Eqs.~(\ref{ar:ERM}) to change variables 
from $a$ and $r_s$ to $\nu$ and $g$ and if we use the
adiabatic relations in Eqs.~(\ref{adiabatic-ERM})
to express the derivatives in terms of matrix elements,
we obtain the virial theorem in Eq.~(\ref{virial-ERM}).

\item
{\bf Inelastic 2-body loss rate}.
The rate at which low-energy atoms and molecules are depleted by 
transitions to pairs of atoms in lower spin states
can be obtained from Eq.~(\ref{Gamma-RM}) by setting 
${\rm Im}\lambda = 0$ and then taking the limit $\lambda \to 0$.
It can be obtained more simply by using the adiabatic relations 
in Eqs.~(\ref{adiabatic-ERM}) to determine the change in the energy
due to small imaginary changes in the parameters $\nu$ and $g$:
%-----------------
\begin{equation}
\Gamma \approx 
-2 \, \int \! d^3R \left( {\rm Im}\,\nu  \langle \phi^\dagger \phi \rangle
+ \frac{{\rm Im}\, g}{m} 
\langle \phi^\dagger \Phi' + \Phi^{'\dagger} \phi \rangle \right) .
\end{equation}
%-----------------

\end{enumerate}

The scattering length in Eq.~(\ref{a:ERM}) can be made arbitrarily 
large by taking the limit $\nu \to 0$ with $g$ fixed.
In this limit, the universality of systems 
with large scattering length requires the Effective Range Model 
to reduce to the Zero-Range Model.  This imposes a constraint 
on the expectation values of the two operators that appear 
in the generalized Tan relations.  The constraint can be derived
by demanding that the
virial theorem in Eq.~(\ref{virial-ERM}) reduce to
the Tan relation in Eq.~(\ref{virial-Tan}) as $\nu \to 0$.
It can also be derived from the constraint for the 
Resonance Model in Eq.~(\ref{constraint-virial:b}) by taking 
the limit $\lambda \to 0$.  They both give the same constraint:
%-----------------
\begin{equation}
g \langle \phi^\dagger \Phi' + \Phi^{'\dagger} \phi \rangle
= -2 m \nu \langle \phi^\dagger \phi \rangle
+ O(\nu^2) .
\label{constraint-ERM}
\end{equation}
%-----------------

%%%%%%%%%%%%%%%%%%%%%%%%%%%%%%%%%%%%%%%%%%%%%%%%%%%%%%%%%%%%%%%%%%%%%%%%%%
\section{Summary}
\label{sec:summary}
%%%%%%%%%%%%%%%%%%%%%%%%%%%%%%%%%%%%%%%%%%%%%%%%%%%%%%%%%%%%%%%%%%%%%%%%%%

The Tan relations enumerated in Section~\ref{sec:Tan}
are universal relations between various properties of an arbitrary system
consisting of fermions with two spin states and a large scattering 
length.  These relations should be satisfied provided the 
energy, temperature, and number density of the system are low enough 
that the interactions between the atoms are accurately described 
by the S-wave phase shift in Eq.~(\ref{delta0:ZRM}),
which depends only on the scattering length $a$.
They are satisfied exactly in the Zero-Range Model, in which the 
phase shift is given by Eq.~(\ref{delta0:ZRM})
up to arbitrarily high energies.  

In Ref.~\cite{Braaten:2008uh}, quantum field theory methods 
were used to derive the Tan relations.  The Zero-Range Model
can be formulated as a local quantum field theory.
Using this formulation, the Tan relations were derived 
by using standard renormalization methods together with the
OPE.  The contact density
operator in this model was identified in Ref.~\cite{Braaten:2008uh}
to be $\lambda_0^2 \psi_1^\dagger \psi_2^\dagger \psi_1 \psi_2$.
One advantage of using quantum field theory methods is that 
it is straightforward to derive the generalizations 
of the Tan relations for any system that can be
formulated as a renormalizable local quantum field theory.
In this paper, we used quantum field theory methods to derive the 
generalized Tan relations for the Resonance Model.
We also derived them for the Effective Range Model
by using the fact that it
can be obtained as a limit of the Resonance Model.

The Resonance Model is defined 
by the S-wave phase shift in Eq.~(\ref{delta0:RM}), which depends 
on three interaction parameters $\nu$, $g$, and $\lambda$.
In Sections~\ref{sec:tail}--\ref{sec:lossrate},
we derived the generalized Tan relations for the Resonance Model.
We identified the contact density operator to be
$\Phi^\dagger \Phi$, where $\Phi$ is the composite operator 
defined in Eq.~(\ref{Phi-def}).
The generalized Tan relations involve expectation values of 
linear combinations of three local operators:  $\Phi^\dagger \Phi$,
$\phi^\dagger \Phi + \Phi^\dagger \phi$, and $\phi^\dagger \phi$.
The scattering length in the Resonance Model can be made 
arbitrarily large by tuning $\nu \to 0$.  The condition that the
generalized Tan relations reduce in this limit to the original 
Tan relations implies the constraints on the 
expectation values of $\Phi^\dagger \Phi$,
$\phi^\dagger \Phi + \Phi^\dagger \phi$, and $\phi^\dagger \phi$
that are given in Eqs.~(\ref{constraint-adiabatic:b}),
(\ref{constraint:b}), and (\ref{constraint:c}).

The Effective Range Model is defined 
by the S-wave phase shift in Eq.~(\ref{delta0:ars}),
which depends on the two parameters $a$ and $r_s<0$.
It can be obtained from the Resonance Model by setting 
$\lambda = 0$, in which case the S-wave phase shift 
is given by Eq.~(\ref{delta0:ERM}) in terms of the
parameters $\nu$ and $g$.
In Section~\ref{sec:ERM}, we enumerated the generalized 
Tan relations for the Effective Range Model.
We identified the contact density operator to be
$g^2 \phi^\dagger \phi$.  The generalized Tan relations involve 
expectation values of linear combinations of 
two local operators:  $\phi^\dagger \phi$
and $\phi^\dagger \Phi' + \Phi^{'\dagger} \phi$,
where $\Phi'$ is the composite operator 
defined in Eq.~(\ref{Phi'-def}).
The scattering length in the Effective Range Model can be made 
arbitrarily large by tuning $\nu \to 0$,
which implies $|a| \gg |r_s|$.  The condition that the
generalized Tan relations reduce in this limit to the original 
Tan relations implies the constraint on the 
expectation values of $\phi^\dagger \phi$
and $\phi^\dagger \Phi' + \Phi^{'\dagger} \phi$
that is given in Eq.~(\ref{constraint-ERM}).

The Tan relations can be tested experimentally by using cold 
trapped atoms near a Feshbach resonance, as discussed at the 
end of Section~\ref{sec:Tan}.  Measurements of the 
momentum distributions can be used to determine the contact $C$ 
and the sum $\langle T \rangle + \langle I \rangle$
of the kinetic and interaction energies.  Measurements of the 
density profiles can be used to determine the potential energy 
$\langle V \rangle$.  The virial theorem in Eq.~(\ref{virial-Tan}) 
is a nontrivial relation between 
$\langle T \rangle + \langle I \rangle$,
$\langle V \rangle$, and $C$.
The adiabatic relation in Eq.~(\ref{adiabatic-Tan})
gives a constraint on the variations in 
$\langle T \rangle + \langle I \rangle$,
$\langle V \rangle$, and $C$ from changing the scattering length.
Both of these relations can be tested experimentally.
Since the ability to change the scattering length is essential both for 
determining $C$ and $\langle T \rangle + \langle I \rangle$
and for testing the adiabatic relation, 
the Feshbach resonance plays a crucial role.
The Resonance Model provides a more detailed microscopic model 
for atoms near a Feschbach resonance than the Zero-Range Model.
Our results on the generalized Tan relations 
in the Resonance Model should be useful for quantifying
the theoretical errors in experimental tests of the Tan relations.

\begin{acknowledgments}
This research was supported in part by the Department of Energy 
under grants DE-FG02-05ER15715 and DE-FC02-07ER41457 and
by the National Science Foundation under grant PHY-0653312.
\end{acknowledgments}

\begin{appendix}
%%%%%%%%%%%%%%%%%%%%%%%%%%%%%%%%%%%%%%%%%%%%%%%%%%%%%%%%%%%%%%
\section{Diagrammatic Calculations}
\label{app:diagrams}

\subsection{Feynman rules}

The results for the matrix elements given in Sections~\ref{sec:tail}
and \ref{sec:pairdensity} can be calculated relatively easily by 
applying a set of rules -- known generally in quantum field theory
as Feynman rules.  The calculation of matrix elements
of operators reduces to drawing all relevant Feynman diagrams,
using the Feynman rules to write down mathematical expressions for the
matrix elements, and evaluating any integrals
that appear in those expressions.

\subsubsection{Propagators and vertices}

We first give the Feynman rules for the Resonance Model that are 
required to calculate Green's functions and T-matrix elements.
A Feynman diagram consists of atom lines (represented by single lines)
and molecule lines (represented by double lines) 
connected by vertices of two types. The mathematical expression for a diagram
is obtained by applying the following rules:
\begin{enumerate}

\item {\bf Four-momenta}. 
Assign a four-momentum $(p_0,\bm{p})$,
where $p_0$ is the energy and $\bm{p}$ is the momentum,
to every external line.  The initial and final four-momenta
are constrained by overall conservation of energy and momentum.
The four-momentum of the internal lines are
constrained only by conservation of energy and momentum at each vertex.

\item {\bf Propagator factors}.
For each internal line with four-momentum $(q_0,\qvec)$,
include the factor 
$i/[q_0 - q^2/(2m) + i \epsilon]$ for an atom line and
$i/[q_0 - \nu - q^2/(4m) + i \epsilon]$ for a molecule line.

\item {\bf Vertex factors}.
For each 2-atom--to--2-atom vertex, 
include the factor $-i \lambda_0/m$.
For each 2-atom--to--molecule vertex, 
include the factor $-i g_0/m$.

\item {\bf Loop momenta}.
If there are four-momenta $(k_0,\bm{k})$ that are not 
determined by the four-momenta of the external lines,
integrate over them using the measure
$\int d^3 k d k_0 /(2\pi)^4$. 
The integrals over $k_0$ can be evaluated using contour integration.

\end{enumerate}

The connected Green's function for $N$ atoms to evolve into
$N$ atoms can be expressed as the sum 
of all connected diagrams with $N$ incoming atom lines and $N$ outgoing 
atom lines.  An example is the amplitude
${\cal A}(E)$ for $N=2$ in Eq.~(\ref{A-E:RM}),
which can be expressed as the infinite sum of diagrams generated 
by iterating the integral equation in Fig.~\ref{fig:inteqRM}.
If this amplitude occurs as a subdiagram
of a Feynman diagram, the Feynman rule for the 
subdiagram is $i {\cal A}(E-P^2/(4m))$, where $E$ and $\bm{P}$ 
are the total energy and momentum flowing through the subdiagram.

The T-matrix element for $N$ atoms to scatter 
into $N$ atoms is obtained by setting the energy
for each external atom line equal to the value required by the
nonrelativistic energy-momentum relation:
$p_0= p^2/(2m)$.  The T-matrix elements involving diatomic
molecules are more complicated and will not be discussed here.

\subsubsection{Operator vertices}

To calculate matrix elements of composite operators,
we also need Feynman rules for the operators.
Energy and momentum can flow into and out of the vertices for the operators.
For simplicity, we consider only operators at a fixed time $t=0$.
The operators listed below in Eqs.~(\ref{rules:1})
annihilate an atom and create an atom, 
so the operator vertex has an incoming atom line 
and an outgoing atom line.
If the incoming and outgoing momenta are $\bm{k}$ and $\bm{k}'$, 
the Feynman rules for the operator vertices are
%-----------------
\begin{subequations}
\begin{eqnarray}
\psi_\sigma^\dagger(\bm{R} - \mbox{$\frac12$} \bm{r})
\psi_\sigma(\bm{R} + \mbox{$\frac12$} \bm{r}): &&
\exp [i\bm{k} \cdot (\bm{R} + \mbox{$\frac12$} \bm{r})] 
\exp [- i \bm{k}' \cdot (\bm{R} - \mbox{$\frac12$} \bm{r})] ,
\\
\psi_\sigma^\dagger \psi_\sigma(\bm{R}): &&
\exp [i (\bm{k} - \bm{k}') \cdot \bm{R}] ,
\\
\psi_\sigma^\dagger \nabla^j \psi_\sigma(\bm{R})
	- \nabla^j \psi_\sigma^\dagger \psi_\sigma(\bm{R}): &&
i (\bm{k} + \bm{k}')^j\exp [i (\bm{k} - \bm{k}') \cdot \bm{R}] .
\end{eqnarray}
\label{rules:1}
\end{subequations}
%-----------------
The operators listed below in Eqs.~(\ref{rules:2}) annihilate a pair
of atoms or a molecule and create a pair of atoms or a molecule. 
If the total momenta entering and leaving the vertex are
$\bm{K}$ and $\bm{K}'$, respectively,
the Feynman rules for the operator vertices are
%-----------------
\begin{subequations}
\begin{eqnarray}
\psi_1^\dagger \psi_2^\dagger \psi_1 \psi_2(\bm{R}): 
&& \exp (i (\bm{K}-\bm{K}') \cdot \bm{R}),
\\
\psi_1^\dagger \psi_2^\dagger \phi(\bm{R}): 
&& \exp (i (\bm{K}-\bm{K}') \cdot \bm{R}),
\\ 
\phi^\dagger \psi_1 \psi_2(\bm{R}): 
&& \exp (i (\bm{K}-\bm{K}') \cdot \bm{R}),
\\
\phi^\dagger \phi(\bm{R}): 
&& \exp (i (\bm{K}-\bm{K}') \cdot \bm{R}).
\end{eqnarray}
\label{rules:2}
\end{subequations}
%-----------------
We will consider matrix elements of these operators only 
between states for which the total momenta $\bm{K}$ 
and $\bm{K}'$ are both zero, in which case the Feynman rules 
in Eqs.~(\ref{rules:2}) all reduce to 1.

The matrix element of a composite operator
between initial and final states consisting of atoms
is the sum of all diagrams in which each external atom line 
is connected either to an operator vertex 
or to other external atom lines.

\subsection{Matrix elements between 2-atom scattering states}

In this subsection, we calculate the matrix elements of 
composite operators between 2-atom scattering states.
We denote the state consisting of a single atom 
with momentum $\bm{p}$ and spin $\sigma$ by
$| \bm{p}, \sigma \rangle$. 
A 2-atom scattering state is labelled by the momenta 
and the spins of the two particles:
$| \bm{p}_1, \sigma_1;  \bm{p}_2, \sigma_2 \rangle$.
For simplicity, we will calculate matrix elements only between
2-atom scattering states consisting of two atoms with
different spins, total momentum 0, 
and total energy $E$.  These states,
which we denote by $| \bm{p}, 1; -\bm{p}, 2 \rangle$,
are labelled by a vector $\bm{p}$ whose magnitude is 
$|\bm{p}| = (mE)^{1/2}$.

\subsubsection{Matrix element of 
$\psi_\sigma^\dagger(\bm{R} - \mbox{$\frac12$} \bm{r})
	\psi_\sigma(\bm{R} + \mbox{$\frac12$} \bm{r})$}
\label{sec:me-psipsi-bilocal}

The matrix element of the bilocal operator 
$\psi_\sigma^\dagger(\bm{R} - \mbox{$\frac12$} \bm{r})
	\psi_\sigma(\bm{R} + \mbox{$\frac12$} \bm{r})$
is given by the sum of the four diagrams 
in Fig.~\ref{fig:<psipsibi>}.  The only one of these diagrams 
that involves an integral over the 4-momentum of an atom 
is Fig.~\ref{fig:<psipsibi>}(d).
Using the Feynman rules, the expression for this diagram is
%-----------------
\begin{eqnarray}
&& \langle \bm{p}', 1; -\bm{p}', 2  | 
	\psi_\sigma^\dagger(\bm{R} - \mbox{$\frac12$} \bm{r})
	\psi_\sigma(\bm{R} + \mbox{$\frac12$} \bm{r}) 
	| \bm{p}, 1; -\bm{p}, 2 \rangle \big|_{\ref{fig:<psipsibi>}(d)} 
\nonumber
\\
&& = i {\cal A}^2(E)
\int \frac{d^3qdq_0}{(2 \pi)^4} 
\frac{\exp (i \bm{q} \cdot \bm{r})}
    {[q_0 - q^2/(2m) + i \epsilon]^2 [E - q_0 - q^2/(2m) + i \epsilon]} ,
\label{<psipsibi>:rules}
\end{eqnarray}
%-----------------
where ${\cal A}(E)$ is the amplitude in Eq.~(\ref{A-E:RM})
and $E = p^2/m$.
The integral over $q_0$ can be evaluated using contours.
The resulting momentum integral is given in Eq.~(\ref{intqexp2}).
The final result is
%-----------------
\begin{equation}
\langle \bm{p}', 1; -\bm{p}', 2  | 
	\psi_\sigma^\dagger(\bm{R} - \mbox{$\frac12$} \bm{r})
	\psi_\sigma(\bm{R} + \mbox{$\frac12$} \bm{r}) 
| \bm{p}, 1; -\bm{p}, 2 \rangle \big|_{\ref{fig:<psipsibi>}(d)} = 
\frac{i m^2}{8 \pi p} {\cal A}^2(E) \exp (i p r) .
\label{<psipsibi>}
\end{equation}
%-----------------
Unlike the first three diagrams in Fig.~\ref{fig:<psipsibi>},
this diagram is not an
analytic function of $\bm{r}$ at $\bm{r} = 0$.

\subsubsection{Matrix element of $\psi_\sigma^\dagger \psi_\sigma$}
\label{sec:me-psipsi-local}

The matrix element of the local operator 
$\psi_\sigma^\dagger \psi_\sigma(\bm{R})$
is given by the sum of the four diagrams in Fig.~\ref{fig:<psipsi>}.  
These diagrams are equal to the $\bm{r} \to 0$ limits of the 
corresponding diagrams for the bilocal operator in 
Fig.~\ref{fig:<psipsibi>}.  For the first three diagrams, 
this can be seen easily from the Feynman rules.
This is not as obvious for the fourth diagram, because it 
involves an integral over the 4-momentum of an atom.
Using the Feynman rules, the expression for 
the diagram in Fig.~\ref{fig:<psipsibi>}(d) is
%-----------------
\begin{eqnarray}
&& \langle \bm{p}', 1; -\bm{p}', 2 | \psi_\sigma^\dagger 
	\psi_\sigma(\bm{R}) | \bm{p}, 1; -\bm{p}, 2 \rangle 
	\big|_{\ref{fig:<psipsi>}(d)}  
\nonumber
\\
&& =i {\cal A}^2(E)
\int \frac{d^3qdq_0}{(2 \pi)^4} 
\frac{1}{[q_0 - q^2/(2m) + i \epsilon]^2
       [E - q_0 - q^2/(2m) + i \epsilon]} .
\label{<psi^2>:rules}
\end{eqnarray}
%-----------------
The integral over $q_0$ can be evaluated using contours.
The resulting momentum integral is given in Eq.~(\ref{intq2}).
The final result is
%-----------------
\begin{equation}
\langle \bm{p}', 1; -\bm{p}', 2 | \psi_\sigma^\dagger 
	\psi_\sigma(\bm{R}) | \bm{p}, 1; -\bm{p}, 2 \rangle 
\big|_{\ref{fig:<psipsi>}(d)} = 
\frac{im^2}{8 \pi p} {\cal A}^2(E) .
\label{<psi^2>}
\end{equation}
%-----------------
This result matches the $r^0$ term in the expansion of
Eq.~(\ref{<psipsibi>}) in powers of $r$.
This is consistent with the Wilson coefficient of
$\psi_\sigma^\dagger \psi_\sigma(\bm{R})$
in the OPE in Eq.~(\ref{OPE-RM}) being simply 1.

\subsubsection{Matrix element of $\Phi^\dagger \Phi$}
\label{sec:me-PhiPhi}

The matrix element for 
$\Phi^\dagger \Phi(\bm{R})$ can be represented 
by the sum of the four diagrams in Fig.~\ref{fig:<PhiPhi>},
together with 12 other diagrams in which there is no scattering 
of the two incoming atoms or the two outgoing 
atoms or both.   Using the Feynman rules, the expression for 
the diagram in Fig.~\ref{fig:<PhiPhi>}(a) is
%-----------------
\begin{eqnarray}
&& \langle \bm{p}', 1; -\bm{p}', 2 | \lambda_0^2 \psi_1^\dagger \psi_2^\dagger 
	\psi_1 \psi_2(\bm{R}) | \bm{p}, 1; -\bm{p}, 2 \rangle 
	\big|_{\ref{fig:<PhiPhi>}(a)}  
\nonumber
\\
&& = \lambda_0^2 \left[ - i{\cal A}(E)
\int \frac{d^3qdq_0}{(2 \pi)^4} 
\frac{1}{[q_0 - q^2/(2m) + i \epsilon]
       [E - q_0 - q^2/(2m) + i \epsilon]} \right]^2.
\label{<psi4>:rules}
\end{eqnarray}
%-----------------
The integral over $q_0$ can be evaluated using contours.
The resulting momentum integral is given in Eq.~(\ref{intq1}).
The  result is
%-----------------
\begin{equation}
\langle \bm{p}', 1; -\bm{p}', 2 | \lambda_0^2 \psi_1^\dagger \psi_2^\dagger 
	\psi_1 \psi_2(\bm{R}) | \bm{p}, 1; -\bm{p}, 2 \rangle 
	\big|_{\ref{fig:<PhiPhi>}(a)}  = 
\lambda_0^2 \left[ m {\cal A}(E)
\left( \frac{\Lambda}{2 \pi^2} + \frac{ip}{4 \pi} \right) \right]^2.
\label{<psi4>}
\end{equation}
%-----------------
The expressions for 
the three diagrams obtained from Fig.~\ref{fig:<PhiPhi>}(a) by 
omitting the scattering of the atoms in the initial state 
or final state or both are obtained from Eq.~(\ref{<psi4>})
by replacing one or both of the factors in square brackets by 1.
The expressions for the three diagrams in 
Figs.~\ref{fig:<PhiPhi>}(b--d) are obtained 
by replacing one or both of the factors of $\lambda_0$ 
by $(g_0^2/m)/(E - \nu_0)$.  The sum of all 16 diagrams is
%-----------------
\begin{equation}
\langle \bm{p}', 1; -\bm{p}', 2 | 
\Phi^\dagger \Phi (\bm{R}) | \bm{p}, 1; -\bm{p}, 2 \rangle
= \left( \lambda_0 + \frac{g_0^2/m}{E - \nu_0} \right)^2 
\left[ 1 + m {\cal A}(E) 
\left( \frac{\Lambda}{2 \pi^2}  + \frac{i p}{4 \pi}  \right) \right]^2 .
\label{<Phi2>sum}
\end{equation}
%-----------------
By using the integral equation in Eq.~(\ref{A-E:inteq}),
this can be simplified to
%-----------------
\begin{equation}
\langle \bm{p}', 1; -\bm{p}', 2 | 
\Phi^\dagger \Phi (\bm{R}) | \bm{p}, 1; -\bm{p}, 2 \rangle
= m^2 {\cal A}^2(E) ,
\label{<Phi2>}
\end{equation}
%-----------------
where ${\cal A}(E)$ is the amplitude in Eq.~(\ref{A-E:RM})
and $E = p^2/m$.
This result matches the $r^1$ term in the expansion of
Eq.~(\ref{<psipsibi>}) in powers of $r$ if the Wilson coefficient of
$\Phi^\dagger \Phi(\bm{R})$
in the OPE in Eq.~(\ref{OPE-ZRM}) is $-r/(8 \pi)$.

\subsubsection{Matrix element of 
$\psi_1^\dagger \psi_1(\bm{R} - \mbox{$\frac12$} \bm{r})
	\psi_2^\dagger \psi_2(\bm{R} + \mbox{$\frac12$} \bm{r})$}
\label{sec:me-n1n2}

The matrix element of the bilocal operator 
$\psi_1^\dagger \psi_1(\bm{R} - \mbox{$\frac12$} \bm{r})
	\psi_2^\dagger \psi_2(\bm{R} + \mbox{$\frac12$} \bm{r})$
is given by the sum of the four diagrams 
in Fig.~\ref{fig:<nnbi>}.  The diagram in
Fig.~\ref{fig:<nnbi>}(a) is simply 1.  The two diagrams in
Fig.~\ref{fig:<nnbi>}(b,c) involve an integral over the 4-momentum 
of an atom..  The diagram in Fig.~\ref{fig:<nnbi>}(d)
involves two such integrals.
Using the Feynman rules, the expression for this diagram is
%-----------------
\begin{eqnarray}
&& \langle \bm{p}', 1; -\bm{p}', 2  | 
	\psi_1^\dagger \psi_1(\bm{R} - \mbox{$\frac12$} \bm{r})
	\psi_2^\dagger \psi_2(\bm{R} + \mbox{$\frac12$} \bm{r}) 
	| \bm{p}, 1; -\bm{p}, 2 \rangle \big|_{\ref{fig:<nnbi>}(d)} 
\nonumber
\\
&&  = \left[ -i {\cal A}(E)
\int \frac{d^3qdq_0}{(2 \pi)^4} 
\frac{\exp (i \bm{q} \cdot \bm{r})}{[q_0 - q^2/(2m) + i \epsilon]
       [E - q_0 - q^2/(2m) + i \epsilon]} \right]^2.
\label{<nn>:rules}
\end{eqnarray}
%-----------------
The integral over $q_0$ can be evaluated using contours.
The resulting momentum integral is given in Eq.~(\ref{intqexp1}).
The final result is
%-----------------
\begin{equation}
\langle \bm{p}', 1; -\bm{p}', 2  | 
	\psi_1^\dagger \psi_1(\bm{R} - \mbox{$\frac12$} \bm{r})
	\psi_2^\dagger \psi_2(\bm{R} + \mbox{$\frac12$} \bm{r}) 
	| \bm{p}, 1; -\bm{p}, 2 \rangle \big|_{\ref{fig:<nnbi>}(d)} = 
\left[ \frac{m}{4 \pi r} {\cal A}(E)  \exp (i p r) \right]^2.
\label{<nn>}
\end{equation}
%-----------------
The term proportional to $r^{-2}$ term in the expansion of
Eq.~(\ref{<nn>}) in powers of $r$ can be matched by the matrix element 
of $\Phi^\dagger \Phi (\bm{R})$ in Eq.~(\ref{<Phi2>})
if the Wilson coefficient of this operator is $1/(16 \pi^2 r^2)$,
in accord with Eq.~(\ref{psi4-ope}).

\subsubsection{Momentum Integrals}

We list here the momentum integrals that arise in the calculation 
of matrix elements in the preceding subsections.  The calculations 
of the matrix elements of bilocal operators in 
Sections~\ref{sec:me-psipsi-bilocal} and \ref{sec:me-n1n2}
require the following Fourier transforms:
%-----------------
\begin{subequations}
\begin{eqnarray}
\int \frac{d^3q}{(2 \pi)^3}  
\frac{\exp (i \bm{q} \cdot \bm{r})}{q^2 - p^2 - i \epsilon} &=& 
\frac{1}{4 \pi r} \exp(i p r),
\label{intqexp1}
\\
\int \frac{d^3q}{(2 \pi)^3}  
\frac{\exp (i \bm{q} \cdot \bm{r})}{(q^2 - p^2 - i \epsilon)^2} &=& 
\frac{i}{8 \pi p} \exp(i p r).
\label{intqexp2}
\end{eqnarray}
\label{intqexp}
\end{subequations}
%-----------------
The calculations of the matrix elements of local operators in 
Sections~\ref{sec:me-psipsi-local} and \ref{sec:me-PhiPhi}
require the following integrals:
%-----------------
\begin{subequations}
\begin{eqnarray}
\int \frac{d^3q}{(2 \pi)^3}
\frac{1}{q^2 - p^2 - i \epsilon} &=& 
\frac{\Lambda}{2 \pi^2} + \frac{ip}{4 \pi} ,
\label{intq1}
\\
\int \frac{d^3q}{(2 \pi)^3} 
\frac{1}{(q^2 - p^2 - i \epsilon)^2} &=& 
\frac{i}{8 \pi p} .
\label{intq2}
\end{eqnarray}
\label{intq}
\end{subequations}
%-----------------
The integral in Eq.~(\ref{intq1}) is ultraviolet divergent.
It has been evaluated using an ultraviolet cutoff 
$|\bm{q}| < \Lambda$.

\end{appendix}


\begin{thebibliography}{99}

\bibitem{sps0706}
S.~Giorgini, L.P.~Pitaevskii, and S.~Stringari,
% ``Theory of ultracold Fermi gases,''
 arXiv:0706.3360

%\cite{Lee:2008fa}
\bibitem{Lee:2008fa}
  D.~Lee,
  %``Lattice simulations for few- and many-body systems,''
  arXiv:0804.3501 [nucl-th].
  
  %\cite{Braaten:2004rn}
\bibitem{Braaten:2004rn}
  E.~Braaten and H.-W.~Hammer,
%  ``Universality in Few-body Systems with Large Scattering Length,''
  Phys.\ Rept.\  {\bf 428}, 259 (2006).

\bibitem{Ho04}
Tin-Lun Ho,
% ``Universal Thermodynamics of Degenerate Quantum Gases in the Unitarity Limit'',
Phys.\ Rev.\ Lett.\ {\bf 92}, 090402 (2004).

%\cite{Son:2005rv}
\bibitem{Son:2005rv}
  D.~T.~Son and M.~Wingate,
%   ``General coordinate invariance and conformal invariance in nonrelativistic
%   physics: Unitary Fermi gas,''
  Annals Phys.\  {\bf 321}, 197 (2006).
%  [arXiv:cond-mat/0509786].
  
\bibitem{Tan0505}
Shina Tan,
% ``Energetics of the Fermi gas which has BEC-BCS crossover,''
arXiv:cond-mat/0505200. 
    
\bibitem{Tan0508}
Shina Tan,
% ``Large momentum part of fermions with large scattering length,''
arXiv:cond-mat/0508320.
    
\bibitem{Tan0803}
Shina Tan, arXiv:0803.0841.
    
\bibitem{Tan-private}
Shina Tan, private communication.
    
%\cite{Braaten:2008uh}
\bibitem{Braaten:2008uh}
  E.~Braaten and L.~Platter,
  %``Exact Relations for a Strongly-interacting Fermi Gas from the Operator
  %Product Expansion,''
  Phys.\ Rev.\ Lett.\  {\bf 100}, 205301 (2008).
%  [arXiv:0803.1125 [cond-mat.other]].

%\cite{Wilson:1969}
\bibitem{Wilson:1969}
K.G.~Wilson,
%   ``Non-Lagrangian Models of Current Algebra,''
Phys.\ Rev.\ {\bf 179}, 1499 (1969).
    
%\cite{Kadanoff:1969}
\bibitem{Kadanoff:1969}
L.P.~Kadanoff,
%   ``Operator Algebra and the Determination of Critical Indices,''
Phys.\ Rev.\ Lett.\ {\bf 23}, 1430 (1969).

%\cite{WZ:1972}
\bibitem{WZ:1972}
K.G.~Wilson and W.~Zimmerman,
%``Operator Produce Expansions and Composite Field Operators 
%	in the General Framework of Quantum Field Theory,''
Commun.\ Math.\ Phys.\ {\bf 24}, 87 (1972).
    
\bibitem{Gross-Wilczek}
D.~J.~Gross and F.~Wilczek,
%``Ultraviolet Behavior Of Non-Abelian Gauge Theories,''
Phys.\ Rev.\ Lett.\  {\bf 30}, 1343 (1973);

\bibitem{Politzer}
H.D.~Politzer,
%``Reliable Perturbative Results for Strong Interactions?,''
Phys.\ Rev.\ Lett.\ {\bf 30}, 1346 (1973).


\bibitem{Collins}
J.C.~Collins,
{\it Renormalization: An Introduction to Renormalization, 
the Renormalization Group and the Operator-Product Expansion}
(Cambridge University Press, 1986).

\bibitem{OPE-condmat}
J.~Cardy, 
{\it Scaling and Renormalization in Statistical Physics}
(Cambridge University Press, 1996); 
D.J.~Amit and V. Martin-Mayor,
{\it Field Theory, the Renormalization Group and Critical Phenomena}
(World Scientific, 2005).

%\cite{Nishida:2007pj}
\bibitem{Nishida:2007pj}
  Y.~Nishida and D.~T.~Son,
  %``Nonrelativistic conformal field theories,''
  Phys.\ Rev.\  D {\bf 76}, 086004 (2007).
%  [arXiv:0706.3746 [hep-th]].

%\cite{Mehen:2007dn}
\bibitem{Mehen:2007dn}
  T.~Mehen,
  %``On Non-Relativistic Conformal Field Theory and Trapped Atoms: Virial
  %Theorems and the State-Operator Correspondence in Three Dimensions,''
  arXiv:0712.0867 [cond-mat.other].

\bibitem{BKZ0709}
E.\ Braaten, M.\ Kusunoki, and D.\ Zhang,
% ``Scattering Models for Ultracold Atoms,''
Annals Phys.\ {\bf 323}, 1770 (2008).
%[arXiv:0709.0499].
    
%\cite{Kaplan:1996nv}
\bibitem{Kaplan:1996nv}
  D.B.~Kaplan,
%``More effective field theory for nonrelativistic scattering,''
  Nucl.\ Phys.\ B {\bf 494}, 471 (1997).
%  [arXiv:nucl-th/9610052].

\bibitem{KMCWH02}
S.J.J.M.F.\ Kokkelmans, J.N.\ Milstein, M.L.\ Chiofalo, R.\ Walser,
        and M.J.\ Holland,
%``Resonance superfluidity: Renormalization of
%        resonance scattering theory,''
        Phys.\ Rev.\ A {\bf 65}, 053617 (2002).

%\cite{Phillips:1997xu}
\bibitem{Phillips:1997xu}
D.R.~Phillips, S.R.~Beane and T.D.~Cohen,
%``Nonperturbative regularization and renormalization: Simple examples  from
%    nonrelativistic quantum mechanics,''
  Annals Phys.\  {\bf 263}, 255 (1998).
%  [arXiv:hep-th/9706070].

%\cite{Petrov:2004}
\bibitem{Petrov:2004}
D.S.~Petrov,
%``Three-Boson Problem near a Narrow Feshbach Resonance,''
 Phys.\ Rev.\ Lett.\ {\bf 93}, 143201 (2004).

\bibitem{TKT0504}
J.E.~Thomas, J.~Kinast, and A.~Turlapov,
% ``Virial Theorem and Universality in a Unitary Fermi Gas,''
Phys.\ Rev.\ Lett.\ {\bf 95}, 120402 (2005).
%[arXiv:cond-mat/0503620].

% \bibitem{Son:0707}
% D.T.~Son,
% ``Three comments on the Fermi gas at unitarity in a harmonic trap,''
% arXiv:0707.1851.

\bibitem{Werner:0803}
Felix Werner,
%``Virial theorems for trapped quantum gases,''
arXiv:0803.3277.


\end{thebibliography}
\end{document}